# Energy spectra and thermal properties of diatomic molecules in the presence of magnetic and AB fields with improved Kratzer potential

G.J.Rampho[1], A.N.Ikot[1&2], C.O.Edet[2] and U.S.Okorie[2&3]

[1]Department of Physics, University of South Africa, Florida 1710, Johannesburg-South Africa
[2]Theoretical Physics Group, Department of Physics, University of Port Harcourt, Choba, Nigeria
[3]Department of Physics, Akwa Ibom State University, Ikot Akpaden, P.M.B. 1167, Uyo, Nigeria

**Abstract**
In the present study, the improved screened Kratzer potential (ISKP) is investigated in the presence of external magnetic and Aharanov-Bohm (AB) fields within the framework of non-relativistic quantum mechanics. The Schrodinger equation is solved via the Nikiforov-Uvarov Functional Analysis (NUFA) method and the energy spectra and the corresponding wave function for the ISKP in the presence of external magnetic fields are obtained in a closed form. The obtained energy spectra are used to study three selected diatomic molecules $(H_2, HCl \text{ and } LiH)$. It is observed that the present of the magnetic and AB fields removes the degeneracy for different values of the control parameter. The thermodynamic and magnetic properties of the ISKP in the present of the magnetic and AB fields are also evaluated. The effects of the control potential parameter on the thermodynamic and magnetic properties of the selected diatomic molecules are discussed.

1. Introduction

The study of the Schrödinger wave equation (SWE) [1, 2] in the non-relativistic quantum mechanics provides the necessary information needed to understand the behaviour of any quantum system [3, 4]. Many researchers in recent times have given great attention to the studies of the Schrödinger equation with different potentials [5-13]. One of the recently proposed potential is the improved screened Kratzer potential that was recently proposed by Ikot et al. [14]. The improved screened Kratzer potential is given by [14]

$$V(r) = -4D_e \left( \frac{a}{r} - \frac{b}{r^2} \right) \left( e^{-\frac{(\alpha+\delta)r}{2}} Cosh\left(\frac{\alpha+\delta}{2}\right)r + \frac{\bar{c}}{2} \right) \quad (1)$$

where $a = r_e$ and $b = r_e^2$, $\alpha$ and $\delta$ represent the screening parameter which may the same or different depending on the kind of system under investigation, $\bar{c}$ is a control parameter that takes the values $-1, 0$ and $1$. The improved screened Kratzer potential of equation (1) incorporates many potential models as special cases. For instance, Screened (*coshine)* Kratzer potential is obtained when $\bar{c}=0, \alpha=\delta$ as $V_{SCK}(r) = -4D_e\left(\frac{a}{r} - \frac{b}{2r^2}\right)e^{-\alpha r}\cosh(\alpha r)$. The Screened Kratzer potential is deduced from equation (1) if we set $\bar{c}=-1, \delta=0$ as $V_{SK}(r) = -2D_e\left(\frac{a}{r} - \frac{b}{2r^2}\right)e^{-\alpha r}$. By adjusting the potential parameters as $\bar{c}=-1, \alpha=\delta=0$, the improved Screened Kratzer potential reduces to the Kratzer potential as $V_K(r) = -2D_e\left(\frac{a}{r} - \frac{a}{2r^2}\right)$ and others as reported in Ref. [14]. In recent times there are many researches in the literature that considered the effects of magnetic and Aharanov-Bohm flux fields on quantum systems. For instance, Ghosh and Nath [15] examined the exact solutions of the Dirac equation for pseudoharmonic oscillator with angle-dependent scalar potential and the superposition of two vector potentials with magnetic monopole and AB field. Mbadjoun et al. [16] scrutinized the effects of gravity, external uniform electromagnetic field and Aharonov–Bohm (AB) flux field on the spectral properties of electron quantum dots (QDs) confined by 2D parabolic harmonic oscillator influenced. Similarly, Karayer used the extended Nikiforov–Uvarov method to obtain the solutions of the radial Schrödinger equation in the presence of external magnetic field and Aharonov–Bohm (AB) flux fields [17] with others related studies reported in Refs. [18-24]. Elsaid et al. [25] investigated the magnetization and the magnetic susceptibility of a single electron confined in a two dimensional parabolic quantum ring under the effect of external uniform magnetic field and acceptor impurity. Gumber et al. [26] determined the thermal and magnetic properties of a cylindrical quantum dot in the presence of external electric and magnetic fields. Ikot et al [27] studied the thermodynamic properties of pseudo-harmonic potential in the presence of external magnetic and Aharanov-Bohm fields via superstatistics formalism for some selected diatomic molecules and other many studies by several authors can be found in refs. [28-36]. In the present work, our purpose is to solve the Schrodinger wave equation with the improved screened Kratzer potential model in the presence of magnetic and AB flux fields using the Nikiforov-Uvarov functional analysis (NUFA) method and used the obtained energy to calculate the partition function and other thermodynamic functions such as entropy, mean free energy, specific heat capacity and magnetic susceptibility. We will discuss the effects of the control parameter on the energy spectra and the thermodynamic properties of the system for the three selected diatomic molecules of $H_2, HCl$ and $LiH$.

The paper is as follows. In section 2, we give a review the NU-Functional Analysis (NUFA) method. Section 3 is devoted to the solutions of the 2D Schrödinger equation with the improved screened Kratzer potential and vector potential $\vec{A}$ under the influence of external magnetic and AB flux fields. In section 4, we study the effects of the control parameter on the behavior of thermodynamics properties in the presence of external fields on some selected diatomic molecules. Discussions of results are presented in section 5. Finally, a brief concluding remarks is given in section 6.

## 2. NU-Functional Analysis (NUFA) Method

Using the concepts of NU method [37], parametric NU [38] method and the functional analysis [39] Ikot et al. [40] proposed a simple and elegant method for solving a second order differential equation of the hypergeometric type called Nikiforov-Uvarov-Functional Analysis method(NUFA) method. This method is easy and simple just as the parametric NU method. As it is well-known the NU is used to solve a second-order differential equation of the form [37]

$$\psi_n''(s) + \frac{\tilde{\tau}(s)}{\sigma(s)}\psi_n'(s) + \frac{\tilde{\sigma}(s)}{\sigma^2(s)}\psi_n(s) = 0, \tag{2}$$

Where $\sigma(s)$ and $\tilde{\sigma}(s)$ are polynomials, at most of second degree, and $\tilde{\tau}(s)$ is a first-degree polynomial. Tezcan and Sever [38] latter introduced the parametric form of NU method in the form

$$\psi'' + \frac{\alpha_1 - \alpha_2 s}{s(1-\alpha_3 s)}\psi' + \frac{1}{s^2(1-\alpha_3 s)^2}\left[-\xi_1 s^2 + \xi_2 s - \xi_3\right]\psi(s) = 0 \tag{3}$$

where $\alpha_i$ and $\xi_i (i=1,2,3)$ are all parameters. It can be observed in equation (3) that the differential equation has two singularities at $s \to 0$ and $s \to 1$, thus we take the wave function in the form,

$$\psi(s) = s^\lambda (1-s)^\nu f(s) \tag{4}$$

Substituting equation (4) into equation (3) leads to the following equation,

$$s(1-\alpha s)f''(s)+\left[\alpha_1+2\lambda-(2\lambda\alpha_3+2\nu\alpha_3+\alpha_2)s\right]f'(s)$$

$$-\alpha_3\left(\lambda+\nu+\frac{\alpha_2}{\alpha_3}-1+\sqrt{\left(\frac{\alpha_2}{\alpha_3}-1\right)^2+\frac{\xi_1}{\alpha_3}}\right)\left(\lambda+\nu+\frac{\alpha_2}{\alpha_3^2}-1+\sqrt{\left(\frac{\alpha_2}{\alpha_3}-1\right)^2+\frac{\xi_1}{\alpha_3^2}}\right) \quad (5)$$

$$+\left[\frac{\lambda(\lambda-1)+\alpha_1\lambda-\xi_3}{s}+\frac{\alpha_2\nu-\alpha_1\alpha_3\nu+\nu(\nu-1)\alpha_3-\frac{\xi_1}{\alpha_3}+\xi_2-\xi_3\alpha_3}{(1-\alpha_3 s)}\right]f(s)=0$$

Equation (5) can be reduced to a Gauss hypergeometric equation if and only if the following functions vanished,

$$\lambda(\lambda-1)+\alpha_1\lambda-\xi_3=0 \quad (6)$$

$$\alpha_2\nu-\alpha_1\alpha_3\nu+\nu(\nu-1)\alpha_3-\frac{\xi_1}{\alpha_3}+\xi_2-\xi_3\alpha_3=0 \quad (7)$$

Thus, equation (5) becomes

$$s(1-\alpha_1 s)f''(s)+\left[\alpha_1+2\lambda-(2\lambda\alpha_3+2\nu\alpha_3+\alpha_2)s\right]f'(s)$$

$$-\alpha_3\left(\lambda+\nu+\frac{\alpha_2}{\alpha_3}-1+\sqrt{\left(\frac{\alpha_2}{\alpha_3}-1\right)^2+\frac{\xi_1}{\alpha_3}}\right)\left(\lambda+\nu+\frac{\alpha_2}{\alpha_3^2}-1+\sqrt{\left(\frac{\alpha_2}{\alpha_3}-1\right)^2+\frac{\xi_1}{\alpha_3^2}}\right)f(s)=0 \quad (8)$$

Solving equations (6) and (7) completely give,

$$\lambda=\frac{(1-\alpha_1)}{2}\pm\frac{1}{2}\sqrt{(1-\alpha_1)^2+4\xi_3} \quad (9)$$

$$\nu=\frac{(\alpha_3+\alpha_1\alpha_3-\alpha_2)\pm\sqrt{(\alpha_3+\alpha_1\alpha_3-\alpha_2)^2+4\left(\frac{\xi_1}{\alpha_3}+\alpha_3\xi_3-\xi_2\right)}}{2} \quad (10)$$

Equation (8) is the hypergeometric equation type of the form,

$$x(1-x)f''(x)+\left[c+(a+b+1)x\right]f'(x)-abf(x)=0 \quad (11)$$

Using equations (4), (8) and (11), we obtain the energy equation and the corresponding wave equation respectively for the NUFA method as follows:

$$\lambda^2+2\lambda\left(\nu+\frac{\alpha_2}{\alpha_3}-1+\frac{n}{\sqrt{\alpha_3}}\right)+\left(\nu+\frac{\alpha_2}{\alpha_3}-1+\frac{n}{\sqrt{\alpha_3}}\right)^2-\left(\frac{\alpha_2}{\alpha_3}-1\right)^2-\frac{\xi_1}{\alpha_3^2}=0 \quad (12)$$

$$\psi(s) = Ns^{\frac{(1-\alpha_1)+\sqrt{(1-\alpha_1)^2+4\xi_3}}{2}} (1-\alpha_3 s)^{\frac{(\alpha_3+\alpha_1\alpha_3-\alpha_2)+\sqrt{(\alpha_3+\alpha_1\alpha_3-\alpha_2)^2+4\left(\frac{\xi_1}{\alpha_3^2}+\alpha_3\xi_3-\xi_2\right)}}{2}} {}_2F_1(a,b,c;s) \qquad (13)$$

where $a, b, c$ are given as follows,

$$a = \sqrt{\alpha_3}\left(\lambda + \nu + \frac{\alpha_2}{\alpha_3} - 1 + \sqrt{\left(\frac{\alpha_2}{\alpha_3}-1\right)^2 + \frac{\xi_1}{\alpha_3}}\right) \qquad (14)$$

$$b = \sqrt{\alpha_3}\left(\lambda + \nu + \frac{\alpha_2}{\alpha_3} - 1 - \sqrt{\left(\frac{\alpha_2}{\alpha_3}-1\right)^2 + \frac{\xi_1}{\alpha_3}}\right) \qquad (15)$$

$$c = \alpha_1 + 2\lambda \qquad (16)$$

## 3. Schrödinger equation with Improved screened Kratzer potential with external magnetic and AB fields

The Hamiltonian operator of a particle that is charged and confined to move with the improved screened Kratzer potential under the combined influence of AB and magnetic fields can be written in cylindrical coordinates. Thus, the Schrodinger wave equation can be written as [19, 21]

$$\left(i\hbar\vec{\nabla} - \frac{e}{c}\vec{A}\right)^2 \psi(r,\phi,z) = 2\mu\left[E_{nm} + 4D_e\left(\frac{a}{r} - \frac{b}{r^2}\right)\left(e^{-\frac{(\alpha+\delta)r}{2}} Cosh\left(\frac{\alpha+\delta}{2}\right)r + \frac{\bar{c}}{2}\right)\right]\psi(r,\phi,z), (17)$$

where $E_{nm}$ denotes the energy level, $\mu$ is the effective mass of the system, the vector potential which is denoted by $\vec{A}$ which can be written as a superposition of two terms $\vec{A} = \vec{A}_1 + \vec{A}_2$ having the azimuthal components [41] and external magnetic field with $\vec{\nabla}\times\vec{A}_1 = \vec{B}, \vec{\nabla}\times\vec{A}_2 = 0$, where $\vec{B}$ is the magnetic field. We choose $\vec{A}_1 = \frac{\vec{B}e^{-(\alpha+\delta)r}}{1-e^{-(\alpha+\delta)r}}\hat{\phi}$ and $\vec{A}_2 = \frac{\Phi_{AB}}{2\pi r}\hat{\phi}$ to represents the additional magnetic flux $\phi_{AB}$ created by a solenoid with $\vec{\nabla}\cdot\vec{A}_2 = 0$. The vector potential in full is written in a simple form as;

$$\vec{A} = \left(0, \frac{\vec{B}e^{-(\alpha+\delta)r}}{1-e^{-(\alpha+\delta)r}} + \frac{\Phi_{AB}}{2\pi r}, 0\right) \qquad (18)$$

Let us take a wave function in the cylindrical coordinates as $\psi(r,\phi) = \frac{1}{\sqrt{2\pi r}} e^{im\phi} R_{nm}(r)$, where $m$ denotes the magnetic quantum number. Inserting this wave function and the vector potential of Eq. (18), we arrive at the following second order radial differential equation of the form:

$$R''(r) + \left[ \frac{2\mu E_{nm}}{\hbar^2} - \frac{2\mu}{\hbar^2}\left( -\frac{P_1}{r} + \frac{P_2}{r^2} - \frac{P_3 e^{-(\alpha+\delta)r}}{r} + \frac{P_4 e^{-(\alpha+\delta)r}}{r^2} \right) + \frac{2m\eta}{r\hbar} \frac{\vec{B} e^{(\alpha+\delta)r}}{(1-e^{(\alpha+\delta)r})} - \frac{\eta^2 \vec{B}^2 e^{-2(\alpha+\delta)r}}{\hbar^2 (1-e^{(\alpha+\delta)r})^2} - \frac{\eta^2 \vec{B} \Phi_{AB} e^{-(\alpha+\delta)r}}{\hbar^2 (1-e^{-(\alpha+\delta)r}) \pi r} - \frac{\left[(m+\xi)^2 - \frac{1}{4}\right]}{r^2} \right] R_{nm}(r) = 0 \quad (19b)$$

In order to surmount the centrifugal barrier, we use the Greene and Aldrich approximation [42] written as follows;

$$\frac{1}{r^2} = \frac{(\alpha+\delta)^2}{\left(1-e^{-(\alpha+\delta)r}\right)^2} \tag{19b}$$

$$R''_{nm}(r) + \left[ \frac{2\mu E_{nm}}{\hbar^2} - \frac{2\mu}{\hbar^2}\left( \begin{array}{c} -\frac{(\alpha+\delta)P_1}{(1-e^{-(\alpha+\delta)r})} - \frac{(\alpha+\delta)^2 P_2}{(1-e^{-(\alpha+\delta)r})^2} - \frac{(\alpha+\delta)P_3 e^{-(\alpha+\delta)r}}{(1-e^{-(\alpha+\delta)r})} \\ + \frac{(\alpha+\delta)^2 P_4 e^{-(\alpha+\delta)r}}{(1-e^{-(\alpha+\delta)r})^2} \end{array} \right) + \frac{2m\eta}{\hbar}\left( \frac{(\alpha+\delta)\vec{B} e^{-(\alpha+\delta)r}}{(1-e^{-(\alpha+\delta)r})^2} \right) - \frac{\eta^2 \vec{B}^2 e^{-2(\alpha+\delta)r}}{\hbar^2 (1-e^{-(\alpha+\delta)r})^2} - \frac{\eta^2 (\alpha+\delta)\vec{B}\Phi_{AB} e^{-(\alpha+\delta)r}}{\hbar^2 (1-e^{-(\alpha+\delta)r})^2 \pi} - \frac{\left[(m+\xi)^2 - \frac{1}{4}\right](\alpha+\delta)^2}{(1-e^{-(\alpha+\delta)r})^2} \right] R_{nm}(r) = 0 \quad (20)$$

where $\eta = -\frac{e}{c}$, $\phi_0 = \frac{hc}{e}$ and $\xi = \frac{\Phi_{AB}}{\phi_0}$.

For Mathematical simplicity, let's introduce the following dimensionless notations;

$$P_1 = 2D_e a + 2D_e a\bar{c}, \; P_2 = D_e b + D_e b\bar{c}, \; P_3 = 2D_e a \text{ and } P_4 = D_e b$$

$$-\varepsilon_{nm} = \frac{2\mu E_{nm}}{\hbar^2(\alpha+\delta)^2}, \Delta_1 = \frac{2\mu P_1}{\hbar^2(\alpha+\delta)}, \Delta_2 = \frac{2\mu P_2}{\hbar^2}, \Delta_3 = \frac{2\mu P_3}{\hbar^2(\alpha+\delta)}, \Delta_4 = \frac{2\mu P_4}{\hbar^2} \; z_1 = \frac{2m\eta \vec{B}}{\hbar(\alpha+\delta)},$$

$$z_2 = \frac{\eta^2 \vec{B}^2}{\hbar^2(\alpha+\delta)^2}, z_3 = \frac{\eta^2 \vec{B}\Phi_{AB}}{\hbar^2(\alpha+\delta)\pi} \text{ and } \gamma = (m+\xi)^2 - \frac{1}{4} \tag{21}$$

Now following the NUFA method with the following substitution $s = e^{-(\alpha+\delta)r}$ into Eq. (20), we obtain the following equation:

$$\frac{d^2 R_{nm}(s)}{ds^2} + \frac{1}{s}\frac{dR_{nm}(s)}{ds} + \frac{1}{s^2(1-s)^2}\left[\begin{array}{c}-(\varepsilon_{nm}+\Delta_3+z_2)s^2 + (2\varepsilon_{nm}-\Delta_1+\Delta_3-\Delta_4+z_1-z_3) \\ -(\varepsilon_{nm}-\Delta_1+\Delta_2+\gamma)\end{array}\right]R_{nm}(s) = 0 \tag{22}$$

where

$$\xi_1 = \varepsilon_{nm} + \Delta_3 + z_2$$
$$\xi_2 = 2\varepsilon_{nm} - \Delta_1 + \Delta_3 - \Delta_4 + z_1 - z_3$$
$$\xi_3 = \varepsilon_{nm} - \Delta_1 + \Delta_2 + \gamma \tag{23}$$
$$\alpha_1 = \alpha_2 = \alpha_3 = 1$$

$$\lambda = \sqrt{\varepsilon_{nm} - \Delta_1 + \Delta_2 + \gamma} \tag{25}$$

$$v = \frac{1}{2} + \sqrt{\frac{1}{4} - z_1 + z_2 + z_3 + \Delta_4 + \Delta_2 + \gamma} \tag{26}$$

Substituting Eqs. (23-26) into Eq. (12), we obtain the energy spectra as follows,

$$\varepsilon_{nm} = \Delta_1 - \Delta_2 - \gamma + \frac{1}{4}\left(\frac{\Delta_1 - \Delta_2 + \Delta_3 - \gamma + z_2 - \left(n + \frac{1}{2} + \sqrt{\frac{1}{4} - z_1 + z_2 + z_3 + \Delta_4 + \Delta_2 + \gamma}\right)^2}{\left(n + \frac{1}{2} + \sqrt{\frac{1}{4} - z_1 + z_2 + z_3 + \Delta_4 + \Delta_2 + \gamma}\right)}\right)^2 \tag{27}$$

Hence, with the value of the dimensionless parameters in Eq. (21) and Eq. (27), we obtain the energy spectra for the improved Kratzer potential in the presence of magnetic and AB fields as,

$$E_{nm} = \frac{\hbar^2(\alpha+\delta)^2}{2\mu}\left[(m+\xi)^2 - \frac{1}{4}\right] - (\alpha+\delta)P_1 + P_2(\alpha+\delta)^2$$

$$-\frac{\hbar^2(\alpha+\delta)^2}{8\mu}\left(\frac{\frac{2\mu P_1}{\hbar^2(\alpha+\delta)} - \frac{2\mu P_2}{\hbar^2} + \frac{2\mu P_3}{\hbar^2(\alpha+\delta)} - \left[(m+\xi)^2 - \frac{1}{4}\right] + \frac{\eta^2 \vec{B}^2}{\hbar^2(\alpha+\delta)^2} - (n+\Omega)^2}{(n+\Omega)}\right)^2 \quad (28)$$

Where

$$\Omega = \frac{1}{2} + \sqrt{(m+\xi)^2 - \frac{2m\eta\vec{B}}{\hbar(\alpha+\delta)} + \frac{\eta^2\vec{B}^2}{\hbar^2(\alpha+\delta)^2} + \frac{\eta^2\vec{B}\Phi_{AB}}{\hbar^2(\alpha+\delta)\pi} + \frac{2\mu P_4}{\hbar^2} + \frac{2\mu P_2}{\hbar^2}} \quad ; m = \pm 1, \pm 2, \pm 3..., \quad (29)$$

The corresponding unnormalized wave function is obtain as

$$R_{nm}(s) = N_{nm} s^{\sqrt{\varepsilon_{nm}-\Delta_1+\Delta_2+\gamma}} (1-s)^{\frac{1}{2}+\sqrt{\frac{1}{4}-z_1+z_2+z_3+\Delta_4+\Delta_2+\gamma}} \, _2F_1(a,b,c;s) \quad (30)$$

## 4. Thermal and Magnetic properties of Improved Screened Kratzer potential (SMKP) for $H_2$, $HCl$ and $LiH$ diatomic molecules in AB and Magnetic fields

The partition function can be computed by straightforward summation over all possible vibrational energy levels accessible to the system. With the energy spectra of equation (28), the partition function $Z(\beta)$ of the improved screened Kratzer potential at finite temperature $T$ is obtained as [43];

$$Z(\beta) = \sum_{n=0}^{\infty} e^{-\beta(E_n - E_0)} \quad (31)$$

with $\beta = \frac{1}{k_B T}$ and with $k_B$ is the Boltzmann constant. In order to evaluate the partition function, we adopt the Euler–Maclaurin formula [43] approach since it gives better results than other approaches. The Euler-Maclaurin summation formula is defined as follows [43],

$$\sum_{n=0}^{\infty} f(x) = \frac{1}{2}f(0) + \int_0^{\infty} f(x)dx - \sum_{p=1}^{\infty} \frac{B_{2p}}{(2p)!} f^{(2p-1)}(0) \quad (32)$$

where $B_{2p}$ are the Bernoulli numbers, $f^{(2p-1)}$ is the derivative of order $(2p-1)$. Up to $p=3$, and with $B_2 = \frac{1}{6}$ and $B_4 = -\frac{1}{30}$ the partition function $Z$ is written as

The energy spectra of Eq. (28) can be recast in the form,

$$E_{nm} = Q - \varphi \left( \frac{R - (n+\Omega)^2}{(n+\Omega)} \right)^2 \tag{33}$$

where,

$$Q = \frac{\hbar^2 (\alpha+\delta)^2}{2\mu} \left( (m+\xi)^2 - \frac{1}{4} \right) - (\alpha+\delta) P_1 + P_2 (\alpha+\delta)^2 \tag{34a}$$

$$R = \frac{2\mu P_1}{\hbar^2 (\alpha+\delta)} - \frac{2\mu P_2}{\hbar^2} + \frac{2\mu P_3}{\hbar^2 (\alpha+\delta)} - \left( (m+\xi)^2 - \frac{1}{4} \right) + \frac{\eta^2 \vec{B}^2}{\hbar^2 (\alpha+\delta)^2}, \varphi = \frac{\hbar^2 (\alpha+\delta)^2}{8\mu} \tag{34b}$$

Hence, on substituting Eq. (33) into Eq. (32), we obtain;

$$Z(\beta) = e^{-\beta \left( \varphi \left( \frac{R-(n+\Omega)^2}{(n+\Omega)} \right)^2 - Q \right) \bigg|_{n_{max}}} \sum_{n=0}^{n_{max}} e^{-\beta \left( Q - \varphi \left( \frac{R-(n+\Omega)^2}{(n+\Omega)} \right)^2 \right)}$$

$$= \frac{1}{2} + \int_0^{n_{max}} f(x) dx - \sum_{p=1}^{\infty} \frac{B_{2p}}{(2p)!} f^{(2p-1)}(0) \tag{35}$$

where

$$f(x) = e^{-\beta \left( Q - \varphi \left( \frac{R-(n+\Omega)^2}{(n+\Omega)} \right)^2 \right)} \tag{36}$$

$$n_{max} = -\Omega + \sqrt{Q} \pm \sqrt{Q-R} \tag{37}$$

It should be noted that Eq. (37) represents the maximum quantum number. To evaluate the integral in Eq. (35), we write the integral as follows;

$$I_1 = \int_{\Omega}^{n_{max}+\Omega} e^{-\beta \left( 2\varphi R + Q - \frac{\varphi R^2}{\rho^2} - \varphi \rho^2 \right)} d\rho \tag{38}$$

where we have defined; $\rho = n + \Omega$ and the integral is evaluated in the limits: $\Omega \leq \rho \leq n_{max} + \Omega$.

With this, Eq. (38) can now be evaluated using Mathematica software as follows;

$$I_1 = -\frac{e^{-2\sqrt{-\beta\varphi}\sqrt{-R^2\beta\varphi}+\beta(-Q-2R\varphi)}\sqrt{\pi}}{4\sqrt{-\beta\varphi}}\left(\begin{array}{l} -Erf\left[\frac{\sqrt{-R^2\beta\varphi}}{\Omega}-\sqrt{-\beta\varphi}\Omega\right]+e^{4\sqrt{-\beta\varphi}\sqrt{-R^2\beta\varphi}}Erf\left[\frac{\sqrt{-R^2\beta\varphi}}{\Omega}+\sqrt{-\beta\varphi}\Omega\right] \\ -Erf\left[\sqrt{-\beta\varphi}\Omega+\sqrt{-\beta\varphi}n_{max}-\frac{\sqrt{-R^2\beta\varphi}}{\Omega+n_{max}}\right]- \\ e^{4\sqrt{-\beta\varphi}\sqrt{-R^2\beta\varphi}}Erf\left[\sqrt{-\beta\varphi}\Omega+\sqrt{-\beta\varphi}\,n_{max}+\frac{\sqrt{-R^2\beta\varphi}}{\Omega+n_{max}}\right] \end{array}\right) \quad (39)$$

Thus, the final expression for the partition function $Z(\beta)$ is obtained as follows;

$$Z(\beta) = \frac{1}{180}\left(\begin{array}{l} \frac{e^{-\beta\left(Q-\frac{\varphi(R-\Omega^2)^2}{\Omega^2}\right)}\left(\begin{array}{l}-2R^6\beta^3\varphi^3 + 3R^4\beta^2\varphi^2\Omega^2\left(-3+2\beta\varphi\Omega^2\right)+6R^2\beta\varphi\Omega^4\left(-1+\Omega^2\left(5+\beta\varphi\left(1-\beta\varphi\Omega^2\right)\right)\right)+ \\ \Omega^9\left(90+\beta\varphi\Omega\left(-30+\beta\varphi\left(3+2\beta\varphi\Omega^2\right)\right)\right)\end{array}\right)}{\Omega^2} + \\ \frac{45e^{-2\sqrt{-\beta\varphi}\sqrt{-R^2\beta\varphi}-\beta(Q+2R\varphi)}\sqrt{\pi}\left(\begin{array}{l}Erf\left[\frac{\sqrt{-R^2\beta\varphi}}{\Omega}-\sqrt{-\beta\varphi}\Omega\right]+Erf\left[\sqrt{-\beta\varphi}\Omega+\sqrt{-\beta\varphi}n_{max}-\frac{\sqrt{-R^2\beta\varphi}}{\Omega+n_{max}}\right]- \\ e^{4\sqrt{-\beta\varphi}\sqrt{-R^2\beta\varphi}}\left(\begin{array}{l}-Erf\left[\frac{\sqrt{-R^2\beta\varphi}}{\Omega}+\sqrt{-\beta\varphi}\Omega\right]+ \\ Erf\left[\sqrt{-\beta\varphi}\Omega+\sqrt{-\beta\varphi}\,n_{max}+\frac{\sqrt{-R^2\beta\varphi}}{\Omega+n_{max}}\right]\end{array}\right)\end{array}\right)}{\sqrt{-\beta\varphi}} \end{array}\right) \quad (40)$$

where $Erf$ denotes the usual error function [28]. In what follows, all thermodynamic and magnetic properties of the Improved screened Kratzer potential, such as the free energy $F(\beta)$, the entropy $S(\beta)$, the internal energy $U(\beta)$, the specific heat capacity $C(\beta)$, magnetization $M(\beta)$ and magnetic susceptibility $\chi_m(\beta)$ can be obtained via the partition function of Eq.

(40). These thermodynamic and magnetic functions for the diatomic molecules system can be calculated from the following expressions [43, 44];

$$\begin{aligned}
F(\beta) &= -\frac{1}{\beta}\ln Z(\beta), \\
U(\beta) &= -\frac{d\ln Z(\beta)}{d\beta}, \\
S(\beta) &= \ln Z(\beta) - \beta\frac{d\ln Z(\beta)}{d\beta}, \\
C(\beta) &= \beta^2 \frac{d^2 \ln Z(\beta)}{d\beta^2}, \\
M(\beta) &= \frac{1}{\beta}\left(\frac{1}{Z(\beta)}\right)\left(\frac{\partial}{\partial \vec{B}} Z(\beta)\right), \\
\chi_m(\beta) &= \frac{\partial M(\beta)}{\partial \vec{B}}
\end{aligned} \qquad (41)$$

## 5. Results and Discussion

In the present study, we use the energy spectrum of Eq. (28) to study the energy spectra of the three selected diatomic molecules of $H_2, HCl$ and $LiH$. The spectroscopic parameters of these molecules are given in Table 1 and taken from Ref. [45]. We used the following conversions factors: $\hbar c = 1973.269$ eV $\overset{0}{A}$ and $1 amu = 931.5 \times 10^6 eV\left(\overset{0}{A}\right)^{-1}$ for all computations [45]. Tables (2-4) show energy eigenvalues for the improved screened Kratzer potential model for $H_2$ diatomic molecule under the influence of AB flux and external magnetic fields with various values of magnetic quantum numbers for $\bar{c} = 0, -1$ and 1. Tables (5-7) show energy eigenvalues for the improved screened Kratzer potential model for $HCl$ diatomic molecule under the influence of AB flux and external magnetic fields with various values of magnetic quantum numbers for $\bar{c} = 0, -1$ and 1. Tables (8-10) shows energy eigenvalues for the improved screened Kratzer potential model for $LiH$ diatomic molecule under the influence of AB flux and external magnetic fields with various values of magnetic quantum numbers for $\bar{c} = 0, -1$ and 1. We observe that when both $\vec{B} = \Phi_{AB} = 0$, there is exist some pseudo-degeneracy in the energy spectra of the molecules. By subjecting the system to the single effect of the magnetic field $(\vec{B} \neq 0, \Phi_{AB} = 0)$, the energy eigenvalues is raised and degeneracy is removed but we still notice some quasi-degeneracy. However, when only the AB field is applied $(\vec{B} = 0, \Phi_{AB} \neq 0)$, the degeneracy is removed and the system becomes more bounded. The all-inclusive effect of both fields is stronger than the single effects and consequently, there is a major shift in the bound state energy of the system. On the other hand, the control parameter tends to create an upward shift and making the system to tend to a continuum state as the control parameter $\bar{c}$ is increased. Table 11 shows a comparison of our energy result with what obtains in Ref.[49] for the ground state for the Kratzer potential with $D_e = 400$ and $r_e = 4$. Table 12 shows our numerical result for $H_2$ is compared with the vibrational states obtained via the Morse potential by other authors.

Figures (1-3) show plots of partition function, free energy, entropy, internal energy, specific heat capacity of improved screened Kratzer potential in magnetic and AB fields for the three diatomic molecules, as a function of $\beta$ for $\bar{c} = -1, 0$ and $1$. In fig. 1(a), the partition function increases linearly as $\beta$ is increased for the three diatomic molecules. In fig. 1(b), the free energy increases monotonically with increasing $\beta$ for the three diatomic molecules. The entropy of the three diatomic molecules decreases with increasing $\beta$ as shown in fig. 1(c). The internal energy of the diatomic molecules also decreases monotonically with increasing $\beta$, as shown in fig. 1(d). Fig 1(e) shows that specific heat capacity of the diatomic molecules increases as $\beta$ is increased and this trend is also observed in figs. 2 and 3. But there is an upward shift in the thermal properties as $\bar{c}$ increases. Figures (4-6) show plots of partition function, free energy, entropy, internal energy, specific heat capacity of improved screened Kratzer potential in magnetic and AB fields for various diatomic molecules, all plotted versus $\vec{B}$ for $\bar{c} = -1, 0$ and $1$. In fig. 4(a), the partition function increases with increase in magnetic field. Fig. 4(b), free energy decreases with increase in magnetic field. The entropy decreases with increasing magnetic field as shown in fig. 4(c). The internal energy decreases with increasing $\vec{B}$ in fig. 4(d). Fig 4(e) shows an increasing specific heat with increasing magnetic field. This trend is also observed in figs. 5 and 6 as clearly seen. But there is an upward shift in the value of the thermal properties as $\bar{c}$ increases. Figs. (7-9) show plots of partition function, free energy, entropy, internal energy, specific heat capacity of improved screened Kratzer potential in magnetic and AB fields for various diatomic molecules, all plotted versus $\Phi_{AB}$ for $\bar{c} = -1, 0$ and $1$. In fig. 7(a), the partition function increases with increase in AB field. In fig. 7(b), free energy decreases with increase in AB field. The entropy decreases with increasing AB field as shown in fig. 7(c). The internal energy decreases with increasing AB field in fig. 7(d). Fig 7(e) shows an increasing specific heat with increasing AB field. This trend is also observed in figs. 8 and 9 as clearly seen. But there is an upward shift in the thermal properties as $\bar{c}$ increases. Figs. 10 (a-c) show plots of magnetization of ISKP in magnetic and AB fields for various diatomic molecules, all plotted versus $\beta$ for $\bar{c} = -1, 0$ and $1$. In figs 10 (a-c), the magnetization increases with increasing $\beta$ in the three cases. But there is an upward shift in the magnetization as $\bar{c}$ increases. Figs. 11 (a-c) show plots of magnetization of ISKP in magnetic and AB fields for various diatomic molecules, all plotted versus magnetic field for $\bar{c} = -1, 0$ and $1$. In figs 11 (a-c), the magnetization decreases with increasing magnetic field in the three cases. Figs. 12 (a-c) show plots of magnetization of ISKP in magnetic and AB fields for various diatomic molecules, all plotted versus AB field for $\bar{c} = -1, 0$ and $1$. In figs. 12 (a-c), the magnetization increases with increasing AB field in the three cases. Figs. 13 (a-c) show plots of magnetic susceptibility of ISKP in magnetic and AB fields for various diatomic molecules, all plotted versus $\beta$ for $\bar{c} = -1, 0$ and $1$. In figs. 13 (a-c), the magnetic susceptibility decreases with increasing $\beta$ when $\bar{c} = -1$ but in fig. 13(b), there is a spread out. The $H_2$ molecule susceptibility decreases with increasing $\beta$ while that of $HCl$ and $LiH$ increases monotonically. Figs. 14 (a-c) show plots of magnetic susceptibility of ISKP in magnetic and AB fields for various diatomic molecules, all plotted versus magnetic field for $\bar{c} = -1, 0$ and $1$. In figs. 14 (a-c), the magnetic susceptibility increases with increasing magnetic field. Figs. 15 (a-c) show plots of magnetic susceptibility of ISKP in magnetic and AB fields for various diatomic molecules, all plotted versus AB field for $\bar{c} = -1, 0$ and $1$. In figs. 15 (a-b), the magnetic susceptibility decreases with increasing magnetic field. However, when $\bar{c} = 1$, the magnetic susceptibility increases with increasing magnetic field as shown in fig. 15 (c).

## 6. Conclusions

In this article, we solve the Schrodinger equation in the presence of AB and magnetic fields for the improved screened Kratzer potential using the NUFA method. We obtain the energy spectrum and wave function for the system. We studied the influence of the control parameter on the energy spectra, magnetic and thermal properties of the system for the three selected diatomic molecules. Our research findings are interesting and could be applied in condensed matter physics, atomic physics and chemical physics.

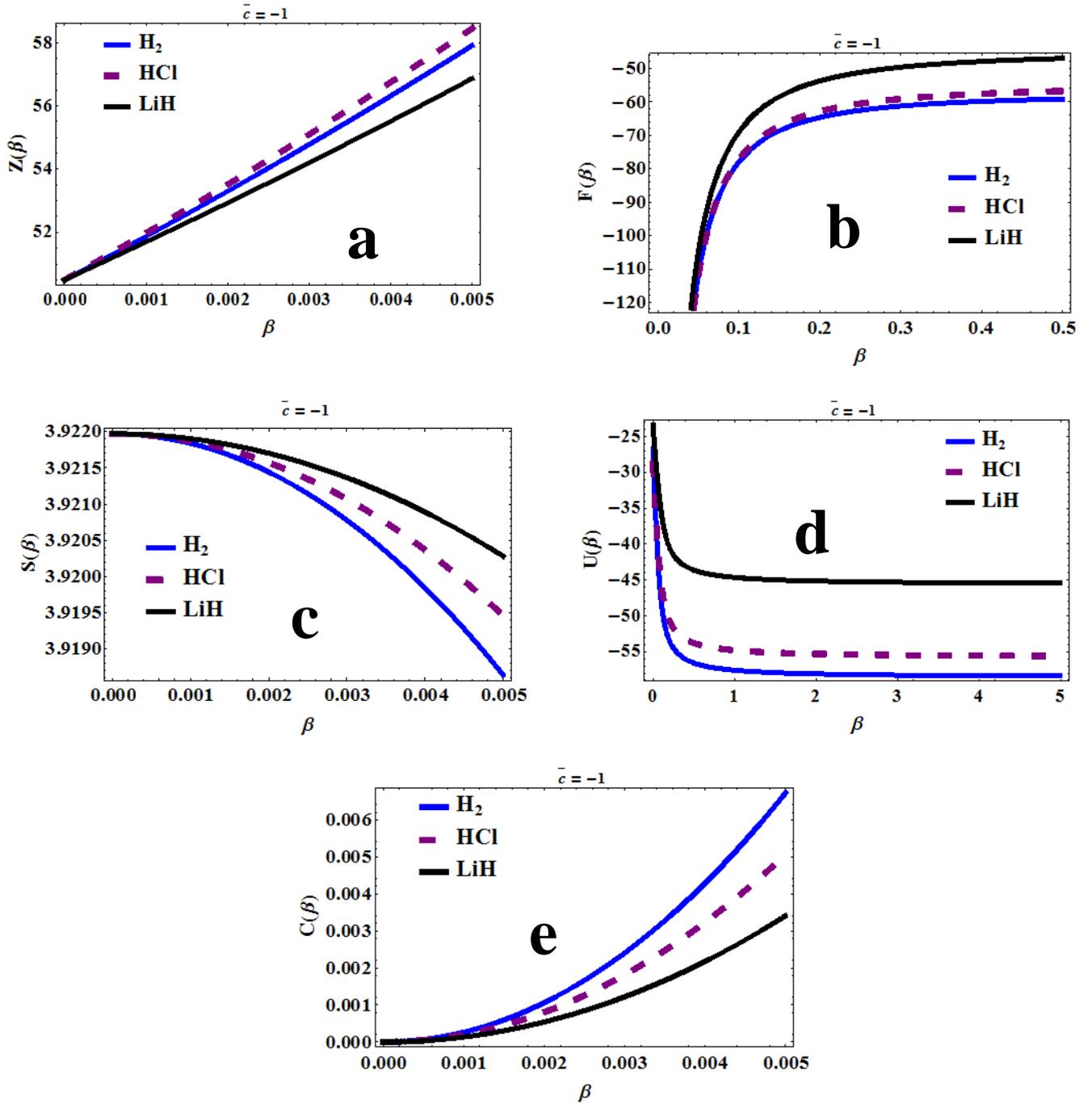

Figure 1: Plots of thermal properties of ISKP in magnetic and AB fields for $\bar{c} = -1$; (a) Partition function of ISKP versus $\beta$ for various diatomic molecules. (b) Free energy of ISKP versus $\beta$ for various diatomic molecules. (c) Entropy of ISKP in magnetic and AB fields versus $\beta$ for various diatomic molecules. (d) Internal energy ISKP versus $\beta$ for various diatomic molecules. (e) specific heat capacity of ISKP versus $\beta$ for various diatomic molecules.

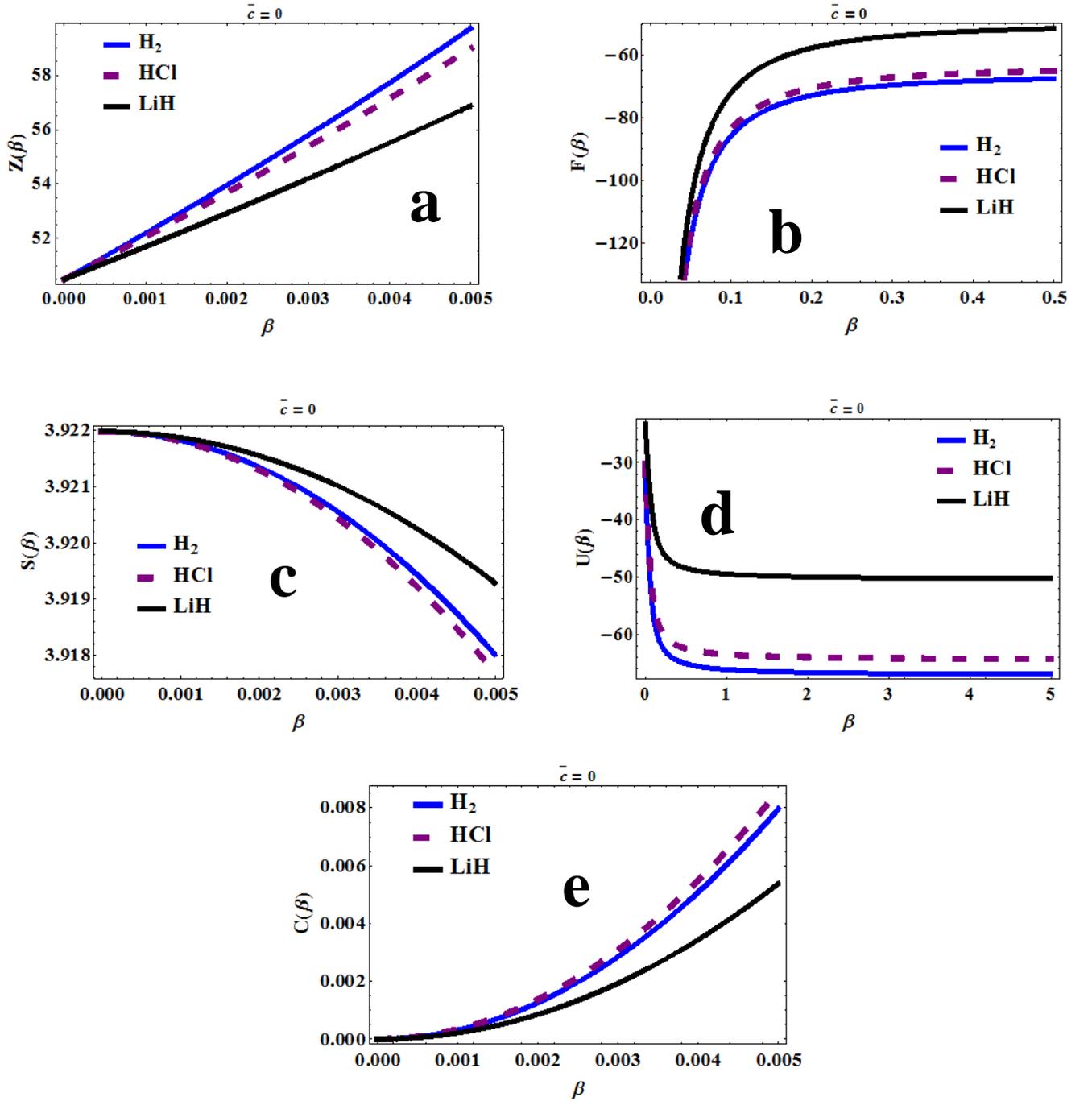

Figure 2: Plots of thermal properties of ISKP in magnetic and AB fields for $\bar{c} = 0$; (a) Partition function of ISKP versus $\beta$ for various diatomic molecules. (b) Free energy of ISKP versus $\beta$ for various diatomic molecules. (c) Entropy of ISKP in magnetic and AB fields versus $\beta$ for various diatomic molecules. (d) Internal energy ISKP versus $\beta$ for various diatomic molecules. (e) specific heat capacity of ISKP versus $\beta$ for various diatomic molecules.

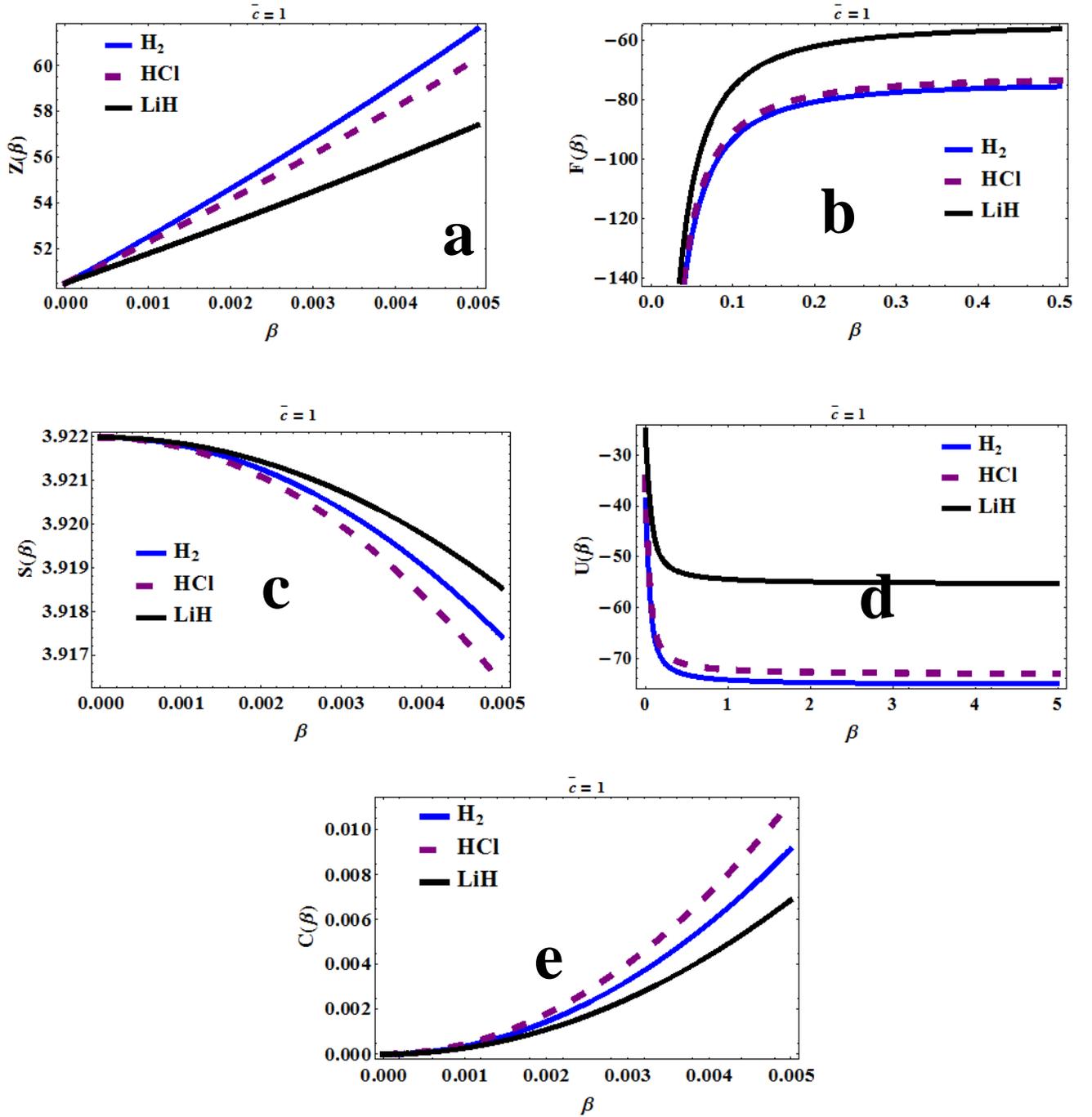

Figure 3: Plots of thermal properties of ISKP in magnetic and AB fields for $\bar{c}=1$; (a) Partition function of ISKP versus $\beta$ for various diatomic molecules. (b) Free energy of ISKP versus $\beta$ for various diatomic molecules. (c) Entropy of ISKP in magnetic and AB fields versus $\beta$ for various diatomic molecules. (d) Internal energy ISKP versus $\beta$ for various diatomic molecules. (e) specific heat capacity of ISKP versus $\beta$ for various diatomic molecules.

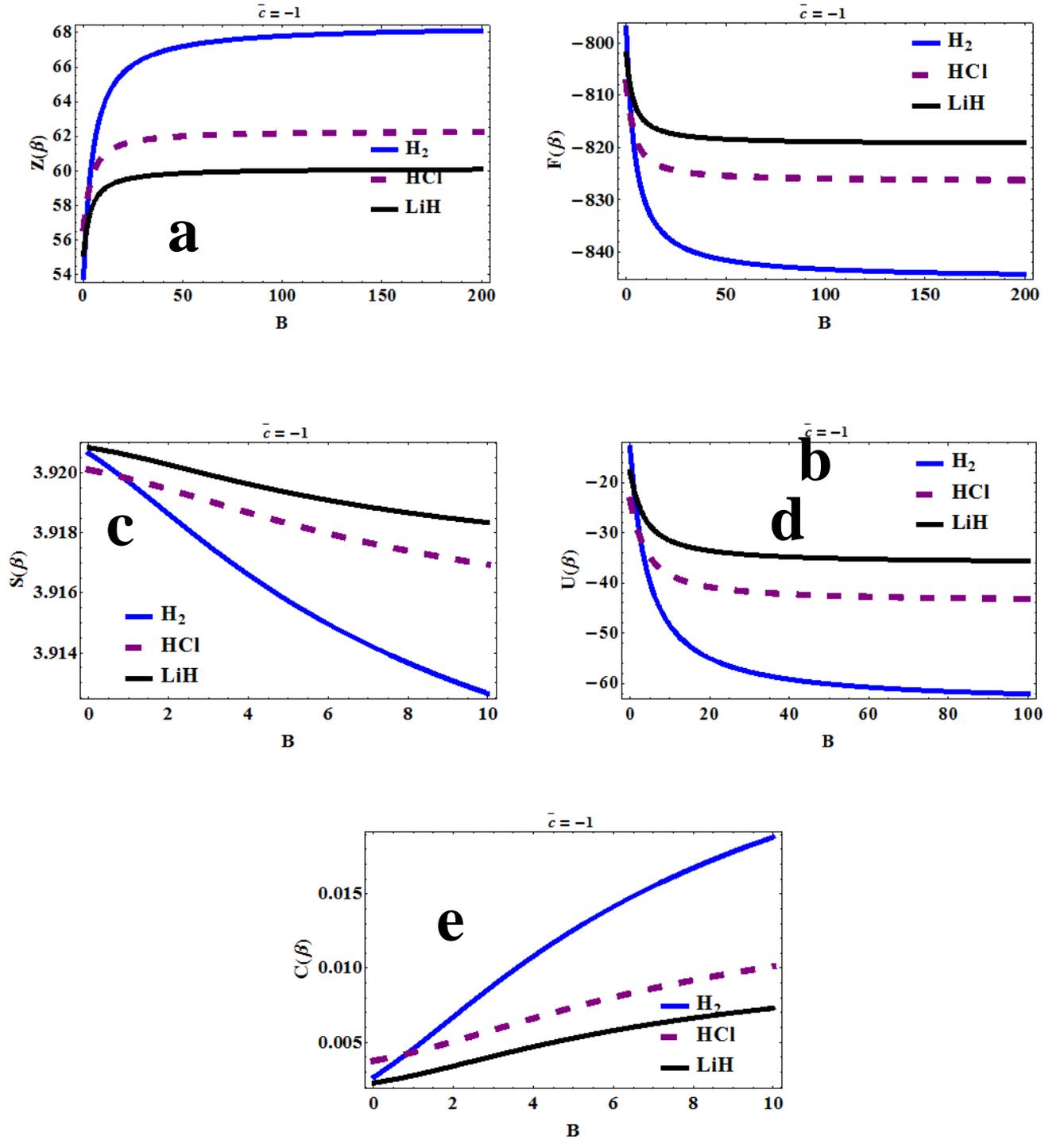

Figure 4: Plots of thermal properties of ISKP in magnetic and AB fields for $\bar{c} = -1$; (a) Partition function of ISKP versus $\vec{B}$ for various diatomic molecules. (b) Free energy of ISKP versus $\vec{B}$ for various diatomic molecules. (c) Entropy of ISKP in magnetic and AB fields versus $\vec{B}$ for various diatomic molecules. (d) Internal energy ISKP versus $\vec{B}$ for various diatomic molecules. (e) specific heat capacity of ISKP versus $\vec{B}$ for various diatomic molecule.

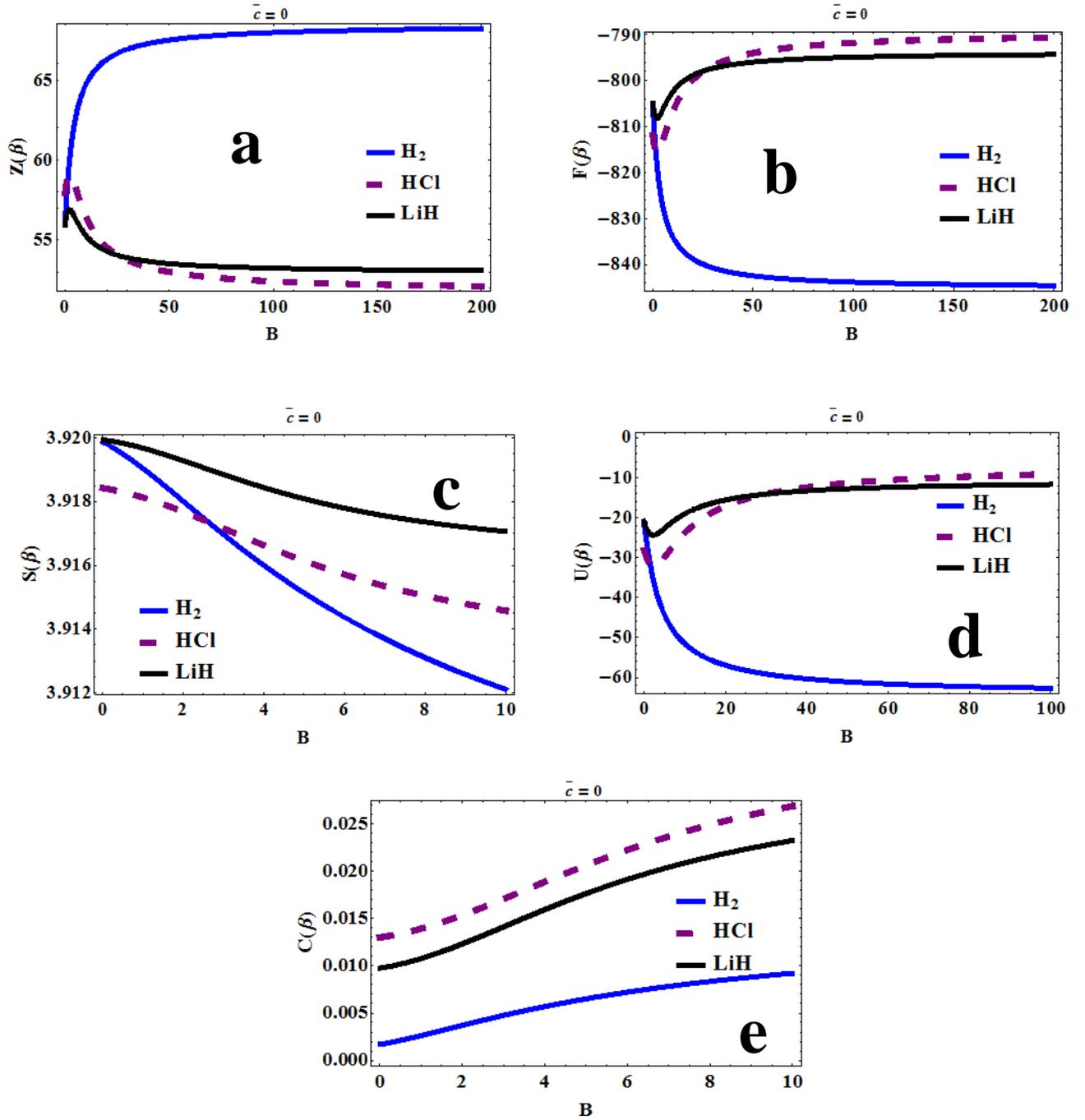

Figure 5: Plots of thermal properties of ISKP in magnetic and AB fields for $\bar{c} = 0$; (a) Partition function of ISKP versus $\vec{B}$ for various diatomic molecules. (b) Free energy of ISKP versus $\vec{B}$ for various diatomic molecules. (c) Entropy of ISKP in magnetic and AB fields versus $\vec{B}$ for various diatomic molecules. (d) Internal energy ISKP versus $\vec{B}$ for various diatomic molecules. (e) specific heat capacity of ISKP versus $\vec{B}$ for various diatomic molecules.

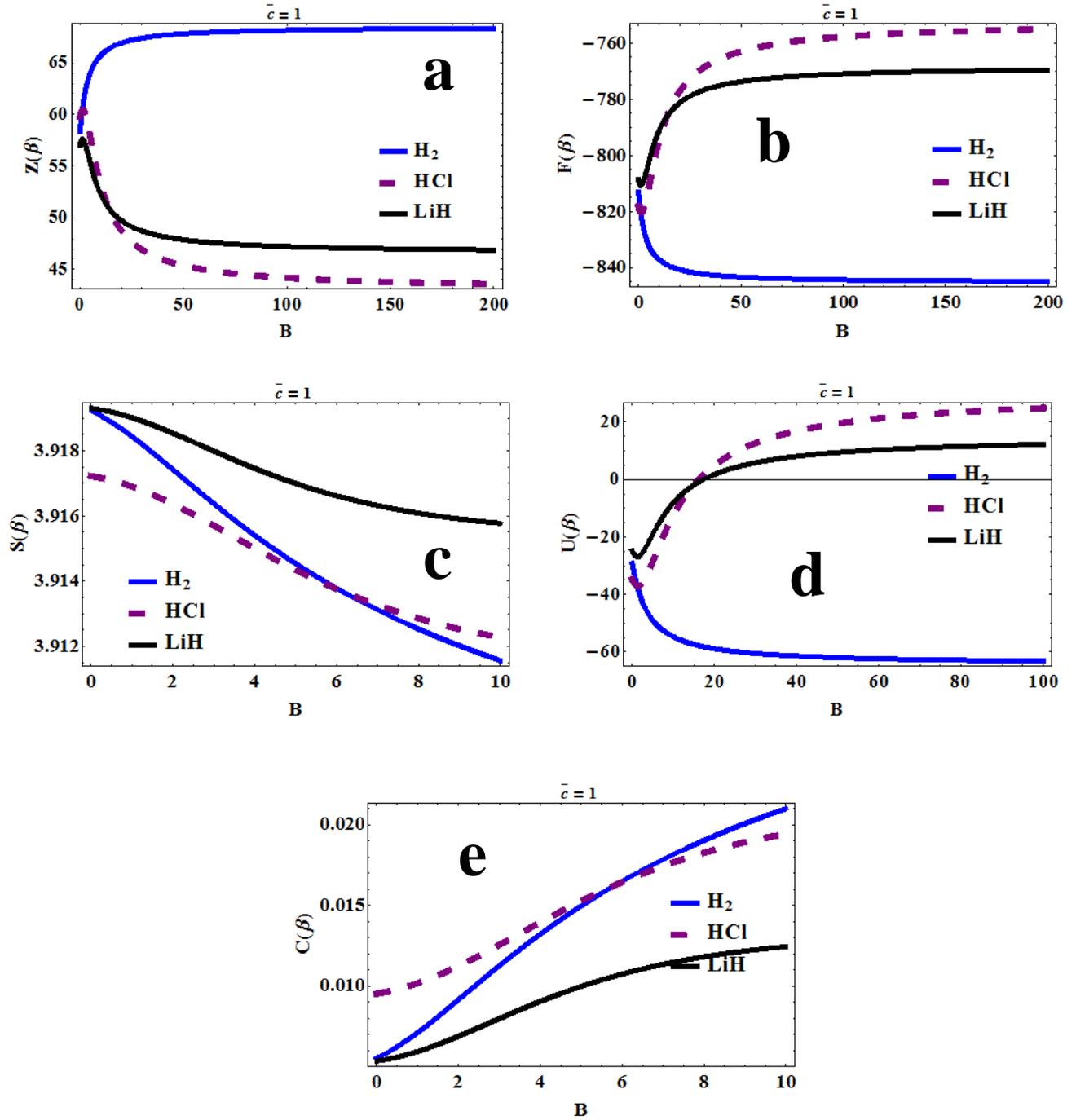

Figure 6: Plots of thermal properties of ISKP in magnetic and AB fields for $\bar{c}=1$; (a) Partition function of ISKP versus $\vec{B}$ for various diatomic molecules. (b) Free energy of ISKP versus $\vec{B}$ for various diatomic molecules. (c) Entropy of ISKP in magnetic and AB fields versus $\vec{B}$ for various diatomic molecules. (d) Internal energy ISKP versus $\vec{B}$ for various diatomic molecules. (e) specific heat capacity of ISKP versus $\vec{B}$ for various diatomic molecules.

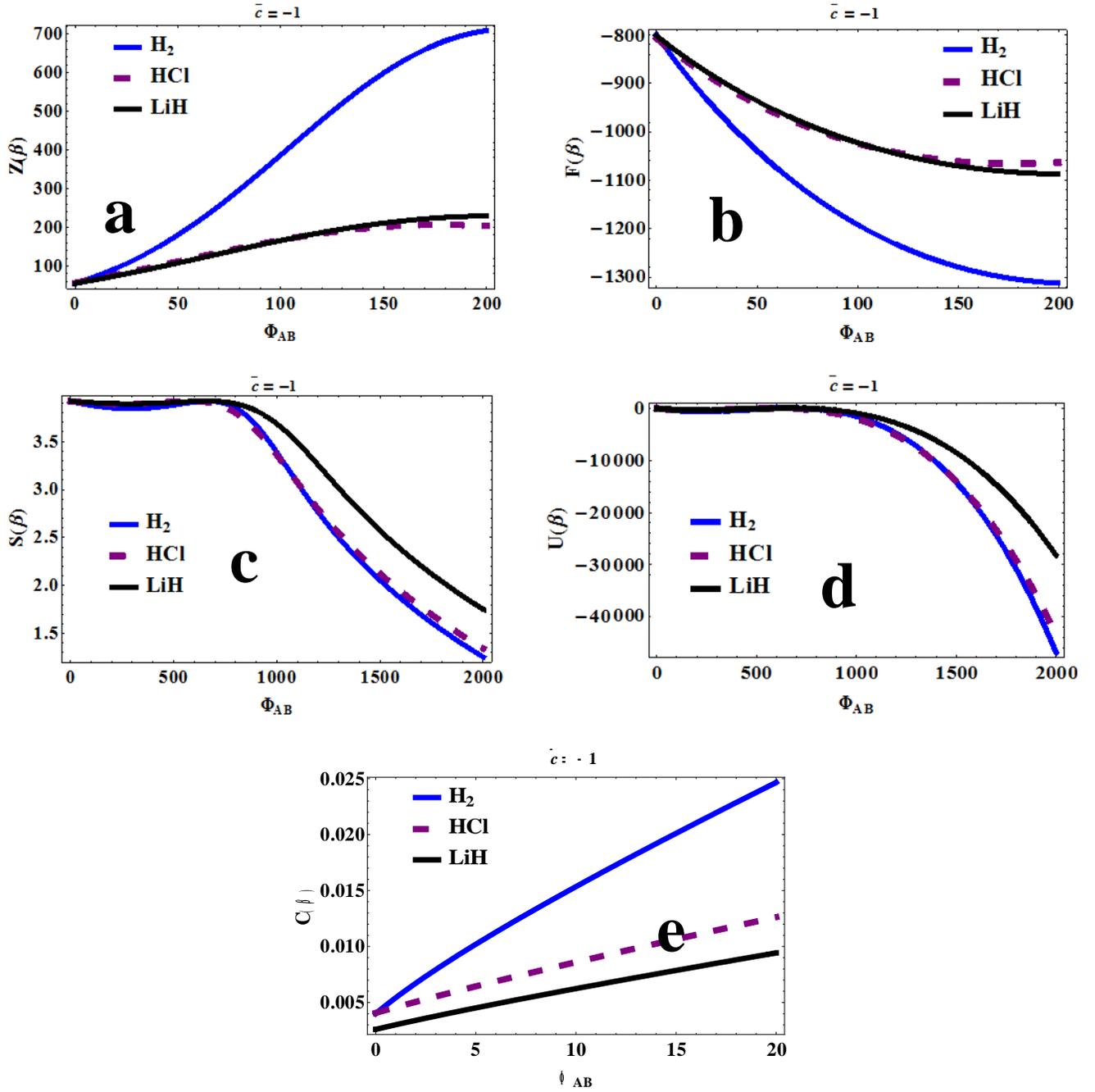

Figure 7: Plots of thermal properties of ISKP in magnetic and AB fields for $\bar{c}=-1$; (a) Partition function of ISKP versus $\Phi_{AB}$ for various diatomic molecules. (b) Free energy of ISKP versus $\Phi_{AB}$ for various diatomic molecules. (c) Entropy of ISKP in magnetic and AB fields versus $\Phi_{AB}$ for various diatomic molecules. (d) Internal energy ISKP versus $\Phi_{AB}$ for various diatomic molecules. (e) specific heat capacity of ISKP versus $\Phi_{AB}$ for various diatomic molecules.

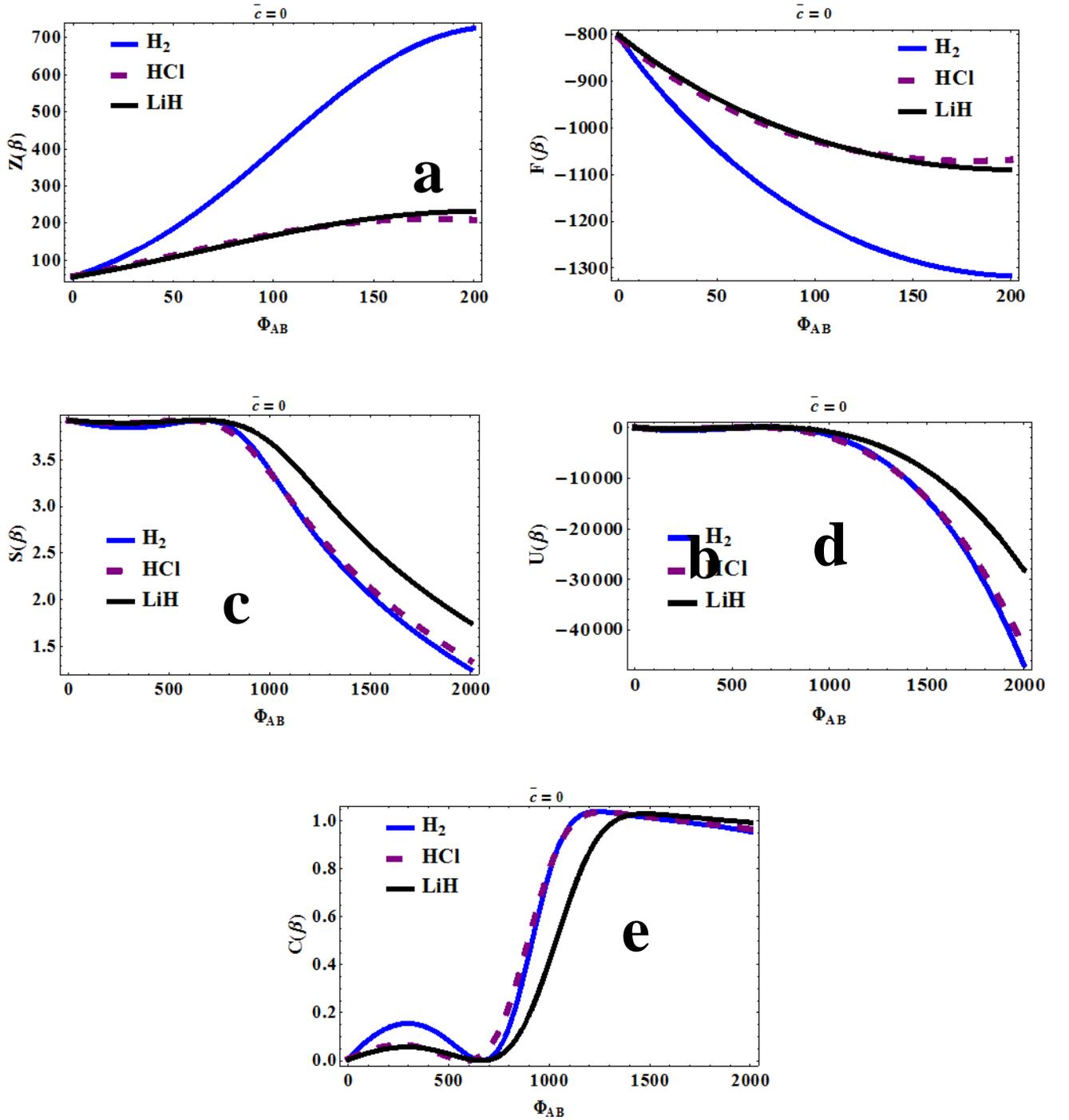

Figure 8: Plots of thermal properties of ISKP in magnetic and AB fields for $\bar{c}=0$; (a) Partition function of ISKP versus $\Phi_{AB}$ for various diatomic molecules. (b) Free energy of ISKP versus $\Phi_{AB}$ for various diatomic molecules. (c) Entropy of ISKP in magnetic and AB fields versus $\Phi_{AB}$ for various diatomic molecules. (d) Internal energy ISKP versus $\Phi_{AB}$ for various diatomic molecules. (e) specific heat capacity of ISKP versus $\Phi_{AB}$ for various diatomic molecules.

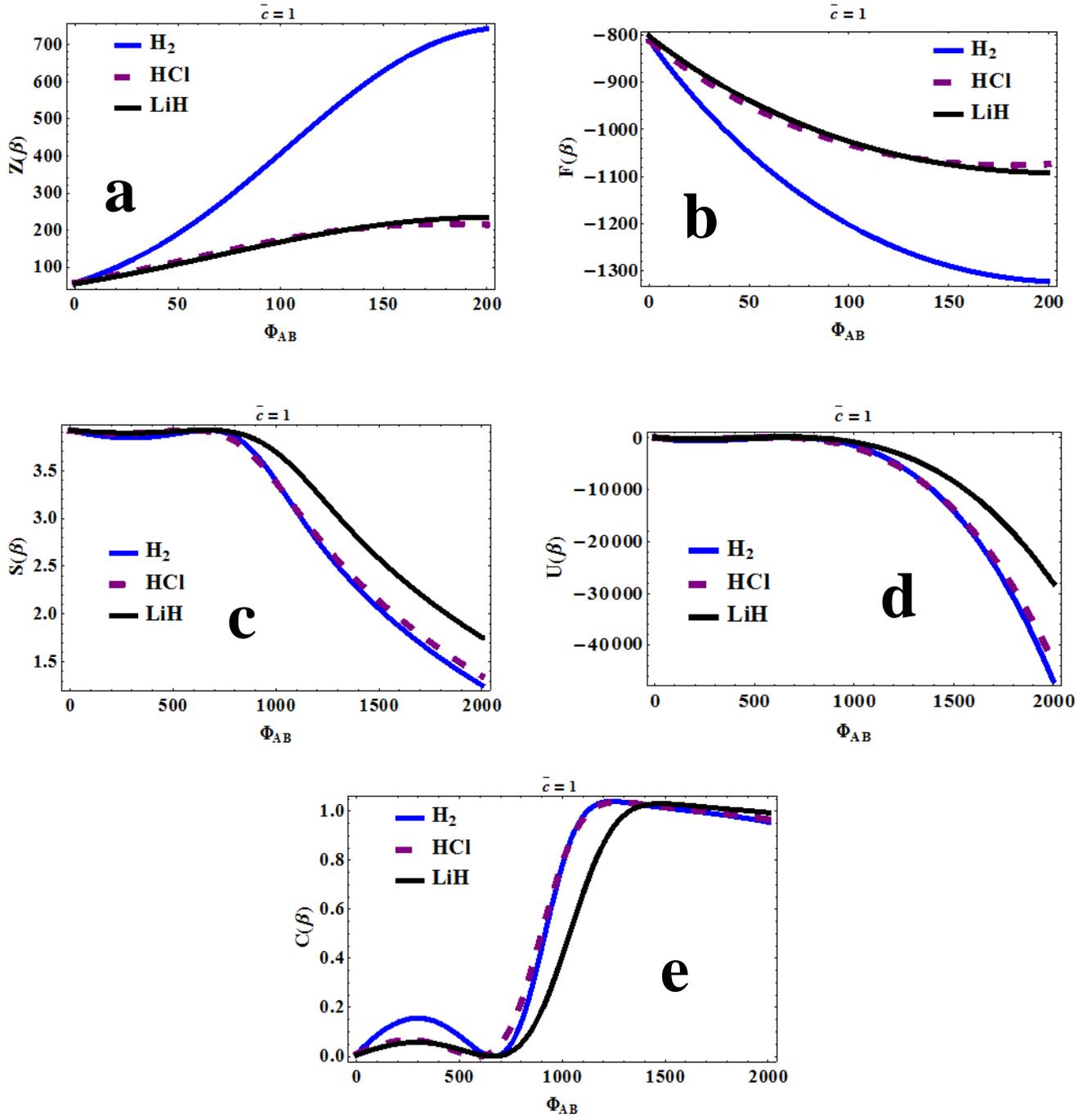

Figure 9: Plots of thermal properties of ISKP in magnetic and AB fields for $\bar{c}=1$; (a) Partition function of ISKP versus $\Phi_{AB}$ for various diatomic molecules. (b) Free energy of ISKP versus $\Phi_{AB}$ for various diatomic molecules. (c) Entropy of ISKP in magnetic and AB fields versus $\Phi_{AB}$ for various diatomic molecules. (d) Internal energy ISKP versus $\Phi_{AB}$ for various diatomic molecules. (e) specific heat capacity of ISKP versus $\Phi_{AB}$ for various diatomic molecules.

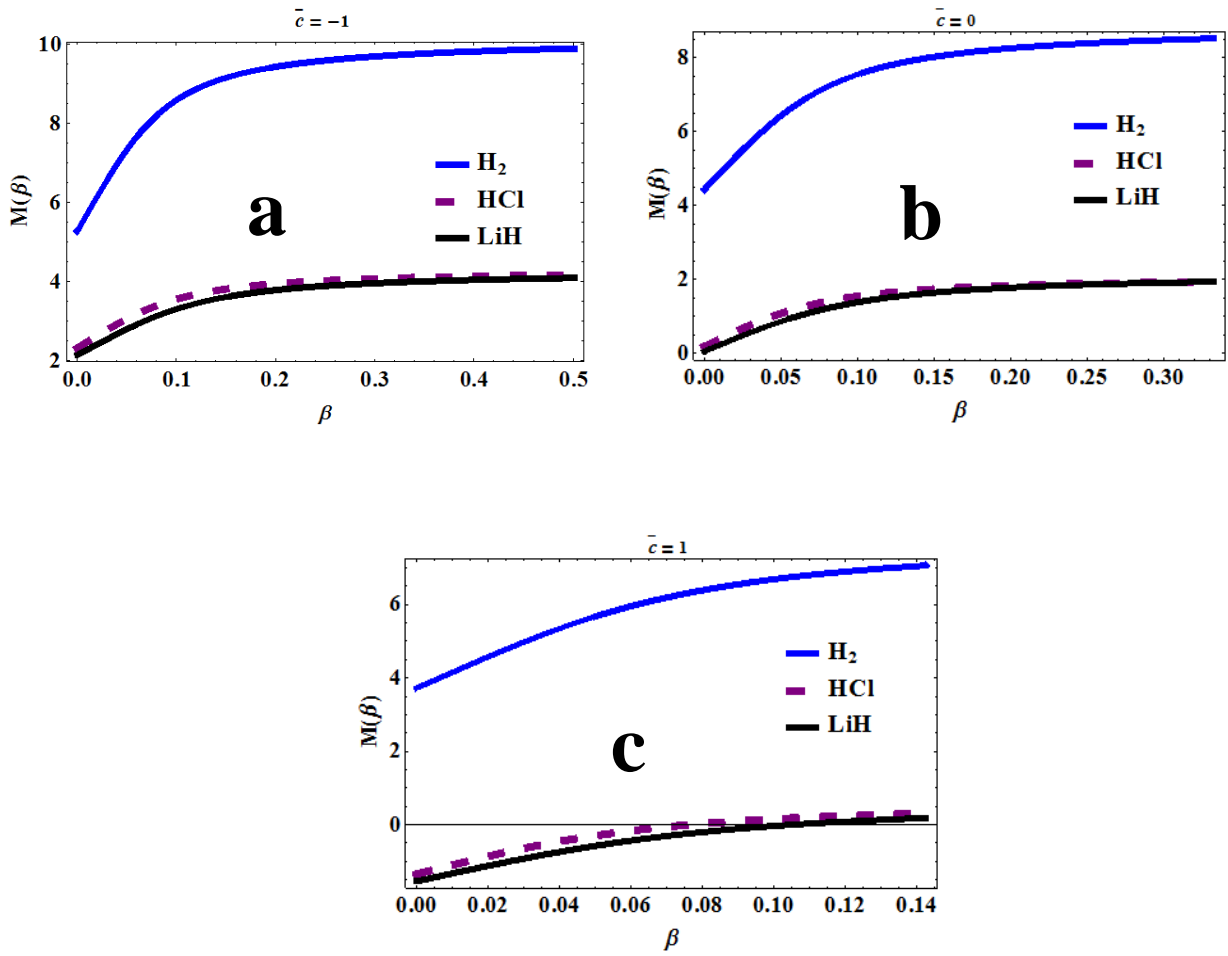

Figure 10: Plots of magnetic properties of ISKP in magnetic and AB fields; (a) Magnetization of ISKP versus $\beta$ for various diatomic molecules for $\bar{c} = -1$. (b) Magnetization of ISKP versus $\beta$ for various diatomic molecules for $\bar{c} = 0$. (c) Magnetization of ISKP versus $\beta$ for various diatomic molecules for $\bar{c} = 1$.

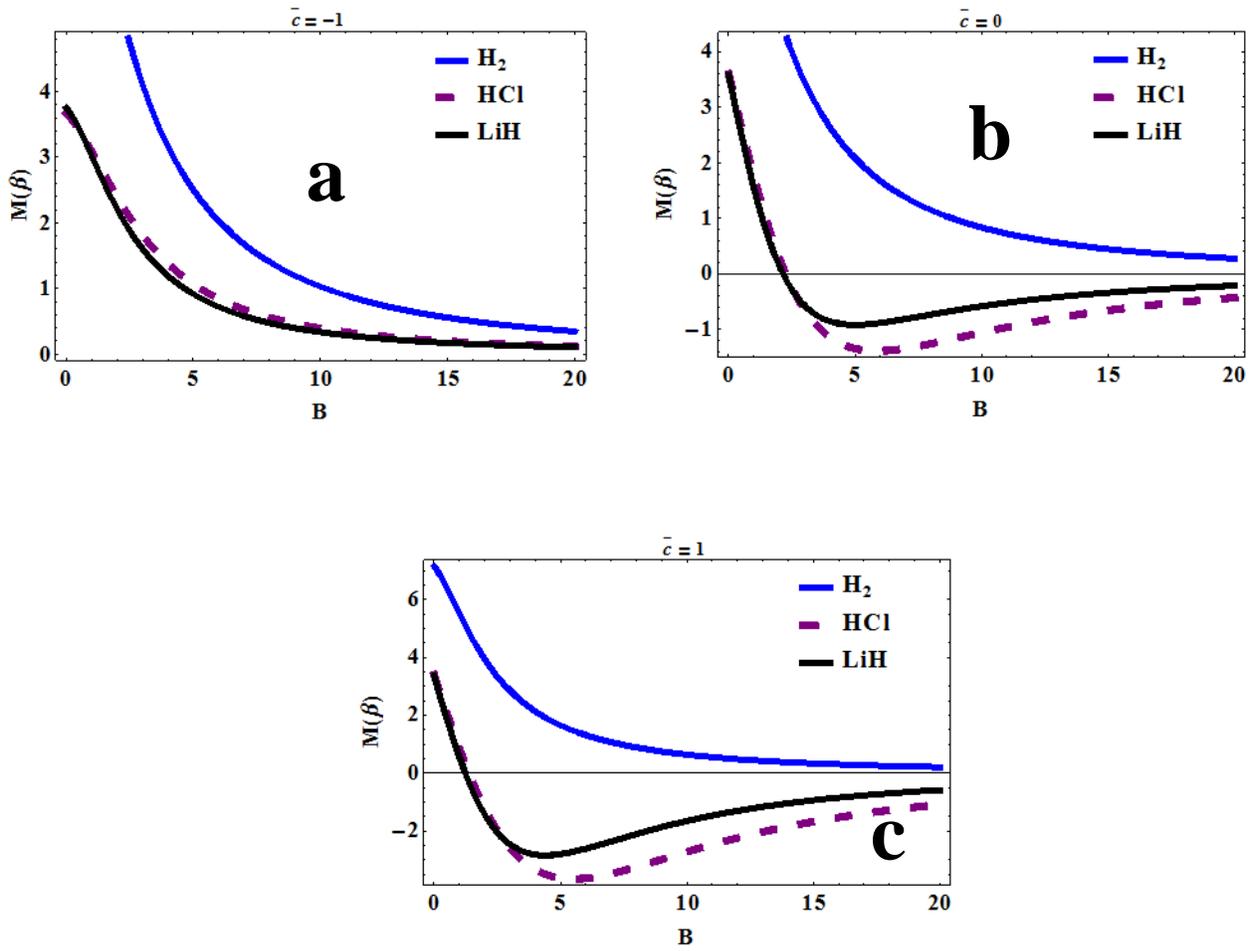

Figure 11: Plots of magnetic properties of ISKP in magnetic and AB fields; (a) Magnetization of ISKP versus $\vec{B}$ for various diatomic molecules for $\bar{c} = -1$. (b) Magnetization of ISKP versus $\vec{B}$ for various diatomic molecules for $\bar{c} = 0$. (c) Magnetization of ISKP versus $\vec{B}$ for various diatomic molecules for $\bar{c} = 1$.

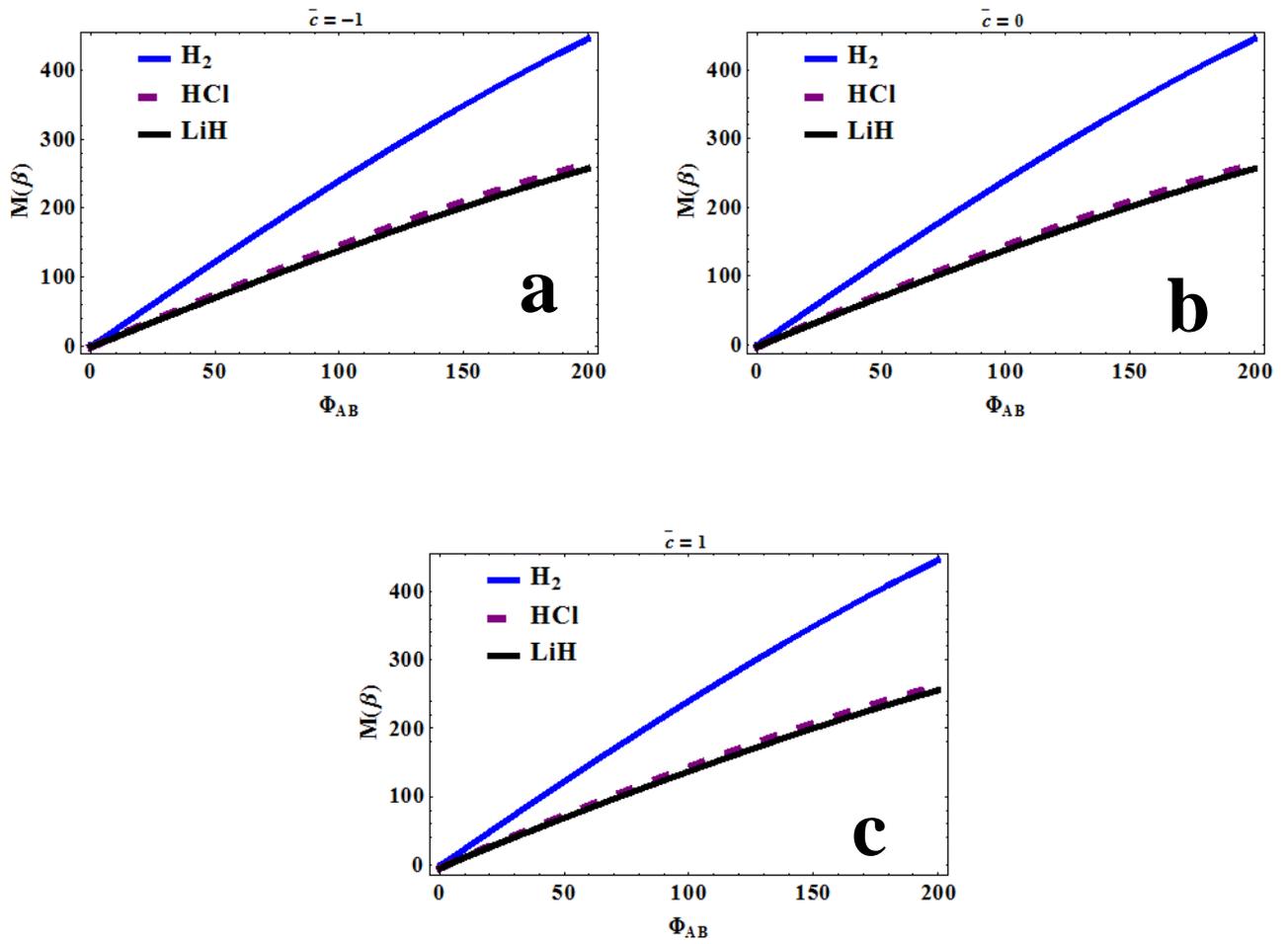

Figure 12: Plots of magnetic properties of ISKP in magnetic and AB fields; (a) Magnetization of ISKP versus $\Phi_{AB}$ for various diatomic molecules for $\bar{c} = -1$. (b) Magnetization of ISKP versus $\Phi_{AB}$ for various diatomic molecules for $\bar{c} = 0$. (c) Magnetization of ISKP versus $\Phi_{AB}$ for various diatomic molecules for $\bar{c} = 1$.

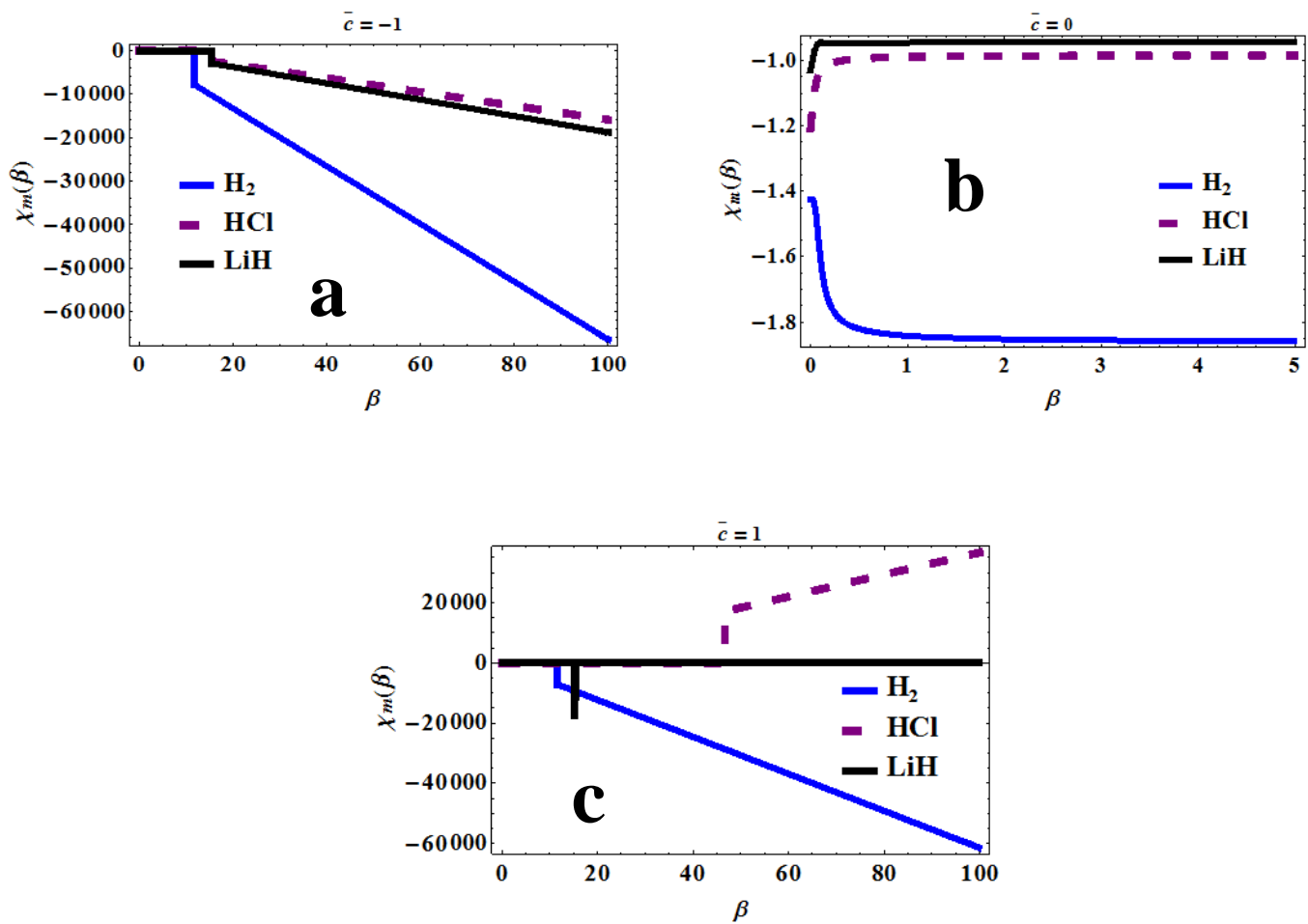

Figure 13: Plots of magnetic properties of ISKP in magnetic and AB fields; (a) Magnetic susceptibility of ISKP versus $\beta$ for various diatomic molecules for $\bar{c}=-1$. (b) Magnetic susceptibility of ISKP versus $\beta$ for various diatomic molecules for $\bar{c}=0$. (c) Magnetic susceptibility of ISKP versus $\beta$ for various diatomic molecules for $\bar{c}=1$.

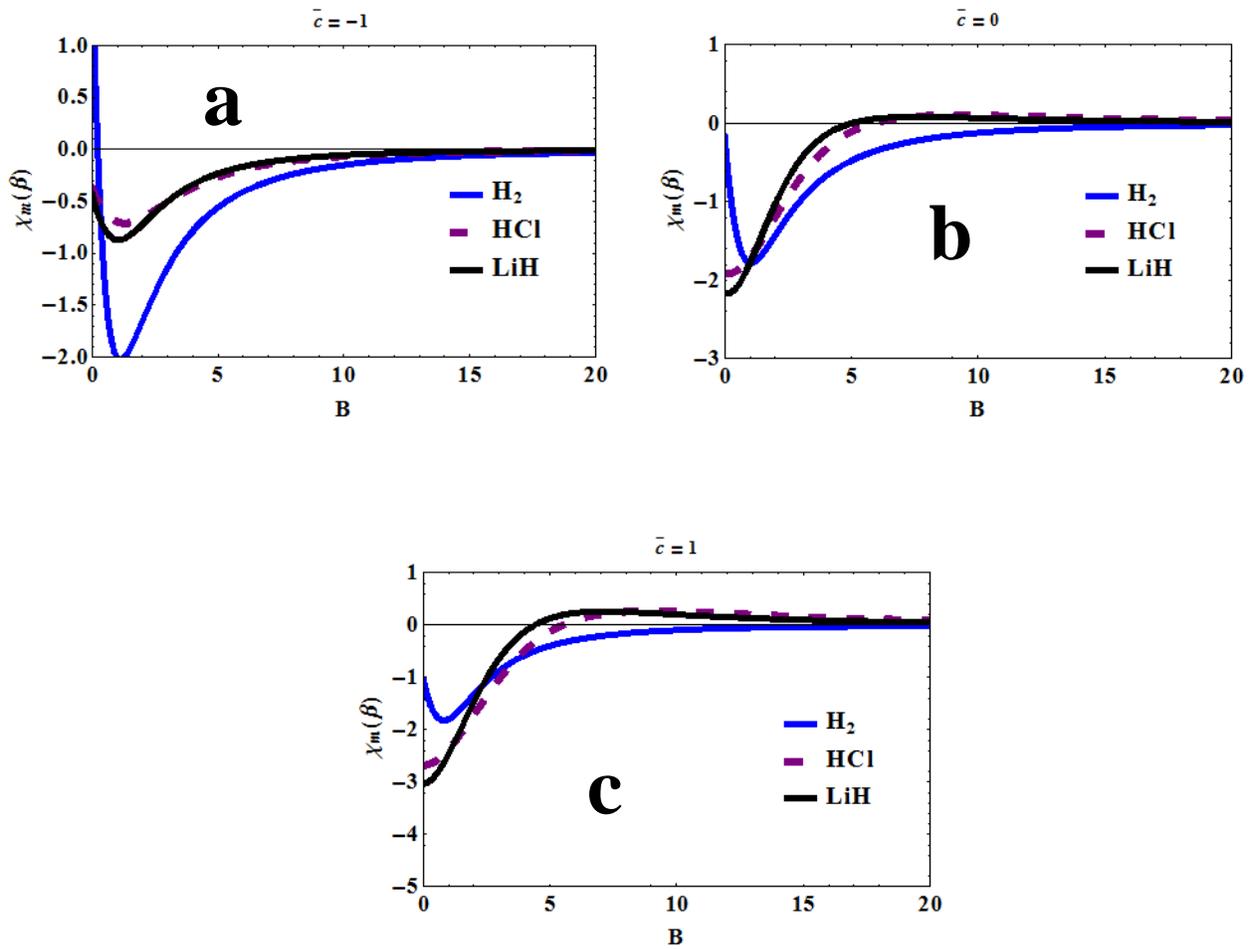

Figure 14: Plots of magnetic properties of ISKP in magnetic and AB fields; (a) Magnetic susceptibility of ISKP versus $\vec{B}$ for various diatomic molecules for $\bar{c} = -1$. (b) Magnetic susceptibility of ISKP versus $\vec{B}$ for various diatomic molecules for $\bar{c} = 0$. (c) Magnetic susceptibility of ISKP versus $\vec{B}$ for various diatomic molecules for $\bar{c} = 1$.

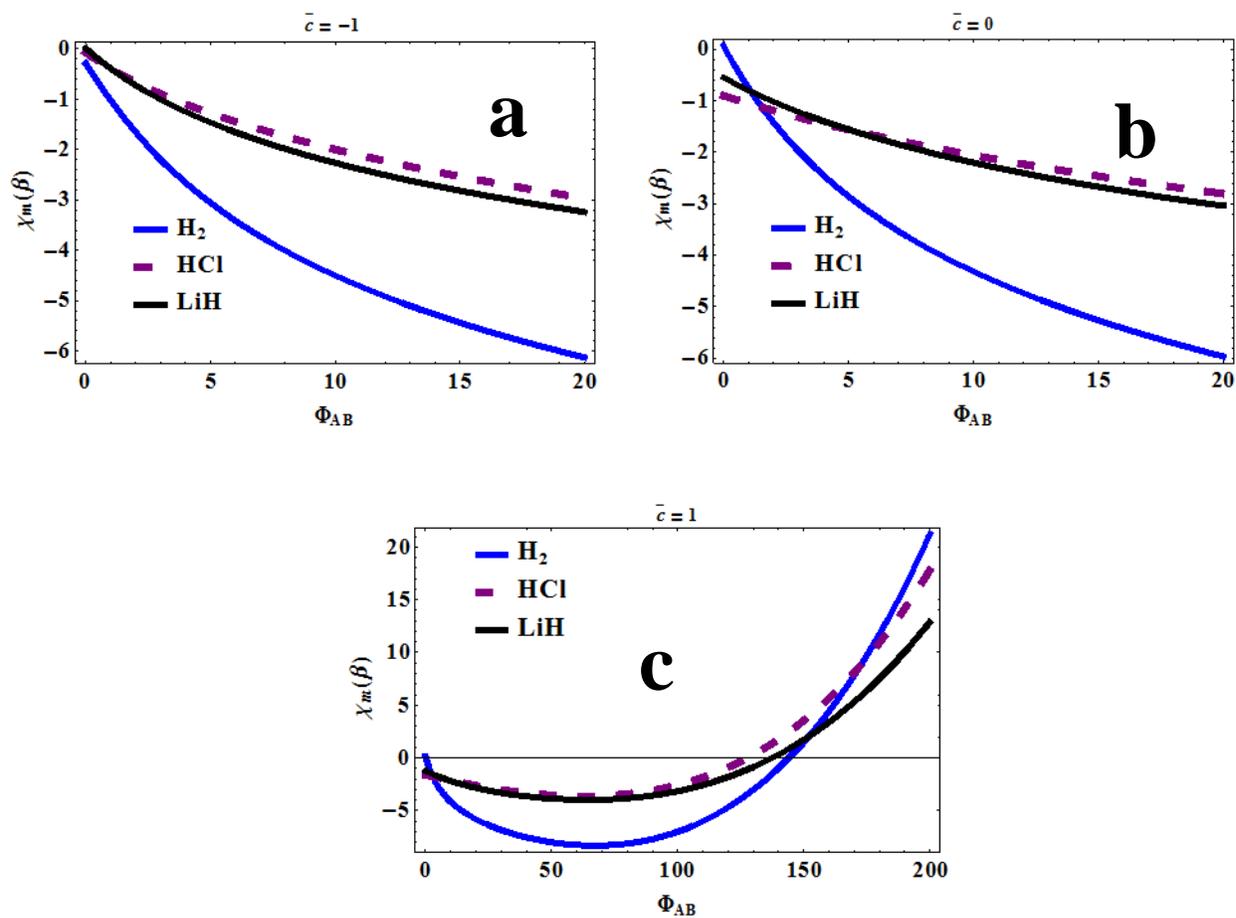

Figure 15: Plots of magnetic properties of ISKP in magnetic and AB fields; (a) Magnetic susceptibility of ISKP versus $\Phi_{AB}$ for various diatomic molecules for $\bar{c}=-1$. (b) Magnetic susceptibility of ISKP versus $\Phi_{AB}$ for various diatomic molecules for $\bar{c}=0$. (c) Magnetic susceptibility of ISKP versus $\Phi_{AB}$ for various diatomic molecules for $\bar{c}=1$.

**Table 1**: Spectroscopic parameters and reduced masses for some diatomic molecules

| Molecules | $D_e(eV)$ | $r_e(\overset{o}{A})$ | $\alpha(\overset{o}{A}{}^{-1})$ | $\mu(amu)$ |
|---|---|---|---|---|
| $HCl$ | 4.619031 | 1.2746 | 1.8677 | 0.980105 |
| $LiH$ | 2.515267 | 1.5956 | 1.128 | 0.880122 |
| $H_2$ | 4.7446 | 0.7416 | 1.9426 | 0.50391 |

**Table2:** Energy values for the improved screened Kratzer potential model for $H_2$ diatomic molecule under the influence of AB flux and external magnetic fields with various values of magnetic quantum numbers for $\bar{c} = -1$.

| | | $\bar{c} = -1$ | | | |
|---|---|---|---|---|---|
| $m$ | $n$ | $\vec{B} = \Phi_{AB} = 0$ | $\vec{B} = 2, \Phi_{AB} = 0$ | $\vec{B} = 0, \Phi_{AB} = 2$ | $\vec{B} = 2, \Phi_{AB} = 2$ |
| 0 | 0 | -0.013053 | -0.013854 | 0.103048 | -3.552120 |
| | 1 | -0.065804 | -0.069578 | 0.043931 | -4.102790 |
| | 2 | -0.169480 | -0.179751 | -0.065444 | -4.684310 |
| | 3 | -0.319135 | -0.341171 | -0.220231 | -5.295650 |
| 1 | 0 | 0.015776 | -0.027578 | 0.248222 | -3.794980 |
| | 1 | -0.039107 | -0.129652 | 0.182814 | -4.367780 |
| | 2 | -0.144694 | -0.283439 | 0.067811 | -4.970660 |
| | 3 | -0.296072 | -0.486006 | -0.092034 | -5.602630 |
| -1 | 0 | 0.016833 | 0.018142 | 0.015776 | -3.272840 |
| | 1 | -0.035913 | 0.010953 | -0.039107 | -3.801170 |
| | 2 | -0.139578 | -0.053657 | -0.144694 | -4.361150 |
| | 3 | -0.289213 | -0.172183 | -0.296072 | -4.951670 |

**Table3:** Energy values for the improved screened Kratzer potential model for $H_2$ diatomic molecule under the influence of AB flux and external magnetic fields with various values of magnetic quantum numbers for $\bar{c} = 0$.

| | | $\bar{c} = 0$ | | | |
|---|---|---|---|---|---|
| $m$ | $n$ | $\vec{B} = \Phi_{AB} = 0$ | $\vec{B} = 2, \Phi_{AB} = 0$ | $\vec{B} = 0, \Phi_{AB} = 2$ | $\vec{B} = 2, \Phi_{AB} = 2$ |
| 0 | 0 | -2.600980 | -1.340270 | -2.523510 | -6.846230 |
| | 1 | -3.021590 | -1.698200 | -2.946390 | -7.547650 |
| | 2 | -3.466830 | -2.094680 | -3.393740 | -8.275340 |
| | 3 | -3.935730 | -2.528000 | -3.864620 | -9.028610 |
| 1 | 0 | -2.584470 | -1.573020 | -2.418360 | -7.157800 |
| | 1 | -3.005820 | -1.959790 | -2.843510 | -7.876480 |
| | 2 | -3.451760 | -2.383820 | -3.292990 | -8.620950 |
| | 3 | -3.921320 | -2.843530 | -3.765840 | -9.390560 |
| -1 | 0 | -2.572960 | -1.073590 | -2.584470 | -6.496540 |
| | 1 | -2.993600 | -1.402100 | -3.005820 | -7.180690 |
| | 2 | -3.438860 | -1.770490 | -3.451760 | -7.891590 |
| | 3 | -3.907770 | -2.176950 | -3.921320 | -8.628520 |

**Table 4:** Energy values for the improved screened Kratzer potential model for $H_2$ diatomic molecule under the influence of AB flux and external magnetic fields with various values of magnetic quantum numbers for $\bar{c} = 1$.

| m | n | $\vec{B} = \Phi_{AB} = 0$ | $\vec{B} = 2, \Phi_{AB} = 0$ | $\vec{B} = 0, \Phi_{AB} = 2$ | $\vec{B} = 2, \Phi_{AB} = 2$ |
|---|---|---|---|---|---|
| | | | $\bar{c} = 1$ | | |
| 0 | 0 | -6.65084 | -3.97026 | -6.58486 | -10.54870 |
| | 1 | -7.24556 | -4.52822 | -7.18090 | -11.37090 |
| | 2 | -7.85971 | -5.11683 | -7.79630 | -12.21650 |
| | 3 | -8.49290 | -5.73507 | -8.43068 | -13.08510 |
| 1 | 0 | -6.63770 | -4.31164 | -6.63770 | -10.15540 |
| | 1 | -7.23286 | -4.89028 | -7.23286 | -10.96340 |
| | 2 | -7.84743 | -5.49886 | -7.84743 | -11.79520 |
| | 3 | -8.48103 | -6.13643 | -8.48103 | -12.65020 |
| -1 | 0 | -6.62423 | -3.59055 | -6.49249 | -10.90350 |
| | 1 | -7.21894 | -4.12773 | -7.08983 | -11.73990 |
| | 2 | -7.83307 | -4.69627 | -7.70645 | -12.59950 |
| | 3 | -8.46623 | -5.29510 | -8.34199 | -13.48170 |

**Table 5:** Energy values for the improved screened Kratzer potential model for $HCl$ diatomic molecule under the influence of AB flux and external magnetic fields with various values of magnetic quantum numbers for $\bar{c} = -1$.

| m | n | $\vec{B} = \Phi_{AB} = 0$ | $\vec{B} = 2, \Phi_{AB} = 0$ | $\vec{B} = 0, \Phi_{AB} = 2$ | $\vec{B} = 2, \Phi_{AB} = 2$ |
|---|---|---|---|---|---|
| | | | $\bar{c} = -1$ | | |
| 0 | 0 | -4.88420 | -3.81271 | -4.82949 | -7.62984 |
| | 1 | -5.37112 | -4.27799 | -5.31739 | -8.24004 |
| | 2 | -5.87571 | -4.76469 | -5.82291 | -8.86802 |
| | 3 | -6.39751 | -5.27222 | -6.34561 | -9.51341 |
| 1 | 0 | -4.87253 | -4.01808 | -4.75515 | -7.81566 |
| | 1 | -5.35977 | -4.49397 | -5.24404 | -8.43432 |
| | 2 | -5.86467 | -4.99099 | -5.75051 | -9.07058 |
| | 3 | -6.38676 | -5.50853 | -6.27411 | -9.72408 |
| -1 | 0 | -4.86447 | -3.57716 | -4.87253 | -7.41507 |
| | 1 | -5.35141 | -4.03196 | -5.35977 | -8.01692 |
| | 2 | -5.85601 | -4.50850 | -5.86467 | -8.63674 |
| | 3 | -6.37782 | -5.00613 | -6.38676 | -9.27416 |

**Table 6:** Energy values for the improved screened Kratzer potential model for *HCl* diatomic molecule under the influence of AB flux and external magnetic fields with various values of magnetic quantum numbers for $\bar{c} = 0$.

| m | n | $\bar{c} = 0$ | | | |
|---|---|---|---|---|---|
| | | $\vec{B} = \Phi_{AB} = 0$ | $\vec{B} = 2, \Phi_{AB} = 0$ | $\vec{B} = 0, \Phi_{AB} = 2$ | $\vec{B} = 2, \Phi_{AB} = 2$ |
| 0 | 0 | -0.445611 | 4.307850 | -0.414125 | -1.084700 |
|   | 1 | -1.339440 | 3.366050 | -1.308080 | -2.088290 |
|   | 2 | -2.243790 | 2.413230 | -2.212550 | -3.102780 |
|   | 3 | -3.158670 | 1.449420 | -3.127560 | -4.128180 |
| 1 | 0 | -0.440385 | 3.978460 | -0.366858 | -1.368560 |
|   | 1 | -1.334270 | 3.032630 | -1.260870 | -2.376000 |
|   | 2 | -2.238690 | 2.075780 | -2.165410 | -3.394350 |
|   | 3 | -3.153640 | 1.107950 | -3.080480 | -4.423580 |
| -1 | 0 | -0.429794 | 4.660870 | -0.440385 | -0.777565 |
|   | 1 | -1.323560 | 3.723070 | -1.334270 | -1.777330 |
|   | 2 | -2.227850 | 2.774250 | -2.238690 | -2.788010 |
|   | 3 | -3.142670 | 1.814420 | -3.153640 | -3.809590 |

**Table 7:** Energy values for the improved screened Kratzer potential model for *HCl* diatomic molecule under the influence of AB flux and external magnetic fields with various values of magnetic quantum numbers for $\bar{c} = 1$.

| m | n | $\bar{c} = 1$ | | | |
|---|---|---|---|---|---|
| | | $\vec{B} = \Phi_{AB} = 0$ | $\vec{B} = 2, \Phi_{AB} = 0$ | $\vec{B} = 0, \Phi_{AB} = 2$ | $\vec{B} = 2, \Phi_{AB} = 2$ |
| 0 | 0 | -2.061300 | 4.692810 | -2.036460 | -0.761398 |
|   | 1 | -3.124230 | 3.552740 | -3.099370 | -1.951810 |
|   | 2 | -4.198530 | 2.402020 | -4.173650 | -3.152850 |
|   | 3 | -5.284160 | 1.240670 | -5.259260 | -4.364520 |
| 1 | 0 | -2.057660 | 4.357560 | -2.057660 | -0.445919 |
|   | 1 | -3.120620 | 3.214270 | -3.120620 | -1.633260 |
|   | 2 | -4.194940 | 2.060330 | -4.194940 | -2.831240 |
|   | 3 | -5.280590 | 0.895753 | -5.280590 | -4.039840 |
| -1 | 0 | -2.047370 | 5.047800 | -1.997710 | -1.057020 |
|   | 1 | -3.110220 | 3.910990 | -3.060510 | -2.250480 |
|   | 2 | -4.184430 | 2.763520 | -4.134680 | -3.454560 |
|   | 3 | -5.269970 | 1.605420 | -5.220190 | -4.669270 |

**Table 8:** Energy values for the improved screened Kratzer potential model for *LiH* diatomic molecule under the influence of AB flux and external magnetic fields with various values of magnetic quantum numbers for $\bar{c} = -1$.

| m | n | $\vec{B} = \Phi_{AB} = 0$ | $\vec{B} = 2, \Phi_{AB} = 0$ | $\vec{B} = 0, \Phi_{AB} = 2$ | $\vec{B} = 2, \Phi_{AB} = 2$ |
|---|---|---|---|---|---|
| 0 | 0 | -3.64587 | -2.45747 | -3.60184 | -6.08325 |
|   | 1 | -4.02055 | -2.80663 | -3.97738 | -6.58099 |
|   | 2 | -4.40904 | -3.17457 | -4.36669 | -7.09355 |
|   | 3 | -4.81098 | -3.56071 | -4.76941 | -7.62059 |
| 1 | 0 | -3.63662 | -2.64604 | -3.54159 | -6.27342 |
|   | 1 | -4.01159 | -3.00647 | -3.91799 | -6.77942 |
|   | 2 | -4.40035 | -3.38533 | -4.30812 | -7.30005 |
|   | 3 | -4.80254 | -3.78205 | -4.71163 | -7.83500 |
| -1 | 0 | -3.62954 | -2.24504 | -3.63662 | -5.86961 |
|   | 1 | -4.00424 | -2.58294 | -4.01159 | -6.35916 |
|   | 2 | -4.39273 | -2.93998 | -4.40035 | -6.86371 |
|   | 3 | -4.79468 | -3.31555 | -4.80254 | -7.38292 |

**Table 9:** Energy values for the improved screened Kratzer potential model for *LiH* diatomic molecule under the influence of AB flux and external magnetic fields with various values of magnetic quantum numbers for $\bar{c} = 0$.

| m | n | $\vec{B} = \Phi_{AB} = 0$ | $\vec{B} = 2, \Phi_{AB} = 0$ | $\vec{B} = 0, \Phi_{AB} = 2$ | $\vec{B} = 2, \Phi_{AB} = 2$ |
|---|---|---|---|---|---|
| 0 | 0 | -0.36375 | 4.85352 | -0.33877 | -0.30138 |
|   | 1 | -1.01847 | 4.14038 | -0.99357 | -1.08083 |
|   | 2 | -1.68203 | 3.41769 | -1.65721 | -1.86961 |
|   | 3 | -2.35444 | 2.68550 | -2.32971 | -2.66770 |
| 1 | 0 | -0.35972 | 4.53924 | -0.30093 | -0.57817 |
|   | 1 | -1.01448 | 3.82170 | -0.95574 | -1.36169 |
|   | 2 | -1.67810 | 3.09463 | -1.61941 | -2.15453 |
|   | 3 | -2.35056 | 2.35807 | -2.29192 | -2.95667 |
| -1 | 0 | -0.35086 | 5.18822 | -0.35972 | -0.00466 |
|   | 1 | -1.00551 | 4.47944 | -1.01448 | -0.78007 |
|   | 2 | -1.66900 | 3.76109 | -1.67810 | -1.56482 |
|   | 3 | -2.34134 | 3.03323 | -2.35056 | -2.35889 |

**Table 10:** Energy values for the improved screened Kratzer potential model for *LiH* diatomic molecule under the influence of AB flux and external magnetic fields with various values of magnetic quantum numbers for $\bar{c} = 1$.

| m | n | $\bar{c} = 1$ | | | |
|---|---|---|---|---|---|
| | | $\vec{B} = \Phi_{AB} = 0$ | $\vec{B} = 2, \Phi_{AB} = 0$ | $\vec{B} = 0, \Phi_{AB} = 2$ | $\vec{B} = 2, \Phi_{AB} = 2$ |
| 0 | 0 | -0.95590 | 6.57443 | -0.93635 | 1.29034 |
| | 1 | -1.72432 | 5.70897 | -1.70473 | 0.37199 |
| | 2 | -2.50261 | 4.83468 | -2.48297 | -0.55518 |
| | 3 | -3.29071 | 3.95157 | -3.27104 | -1.49118 |
| 1 | 0 | -0.95313 | 6.24938 | -0.95313 | 1.59817 |
| | 1 | -1.72157 | 5.38052 | -1.72157 | 0.68304 |
| | 2 | -2.49987 | 4.50282 | -2.49987 | -0.24092 |
| | 3 | -3.28799 | 3.61629 | -3.28799 | -1.17370 |
| -1 | 0 | -0.94464 | 6.91656 | -0.90558 | 0.99964 |
| | 1 | -1.71297 | 6.05454 | -1.67381 | 0.07811 |
| | 2 | -2.49117 | 5.18369 | -2.45192 | -0.85226 |
| | 3 | -3.27918 | 4.30400 | -3.23986 | -1.79145 |

Table 11: Comparison of our energy result with what obtains in Ref.[49] for the ground state for the Kratzer potential with $D_e = 400$ and $r_e = 4$.

| Present | Baoa and Shizgal [49] |
|---|---|
| 9.63436 | 9.63435995 |
| 27.87504 | 27.8750413 |
| 44.85073 | 44.85072948 |
| 60.67575 | 60.67574666 |
| 75.45179 | 75.45178619 |
| 89.26955 | 89.26955046 |

Table 12: Our numerical result for $H_2$ is compared with the vibrational states obtained via the Morse potential by other authors.

| Present | Baoa and Shizgal [49] | Walton et al.[46] | Hunt et. al. [47] | Roy [48] |
|---|---|---|---|---|
| 0.521945198 | 0.5124 | 0.5162 | 0.516 | 0.5137 |
| 0.822531792 | 0.9935 | 1.0032 | 1.0029 | 0.9961 |
| 1.092128962 | 1.443 | 1.4615 | 1.4612 | |
| 1.334854672 | 1.862 | 1.8917 | 1.8912 | |
| 1.554164922 | 2.25 | 2.2937 | 2.2932 | |
| 1.752977462 | 2.606 | 2.6674 | | |
| 1.933769336 | 2.931 | 3.0124 | | |

**References**

[1] W. Greiner, Relativistic Quantum Mechanics: Wave equations (Berlin: Springer) (2000)

[2] L.D. Landau and E.M. Lifshitz, Quantum Mechanics, Non-relativistic Theory (New York: Pergamon) (1977)

[3] L.I. Schiff, Quantum Mechanics (New York: McGraw Hill) (1995)

[4] P.A. Dirac,The Principles of Quantum Mechanics (USA: Oxford University Press) (1958)

[5] C.O. Edet, K.O. Okorie, H. Louis and N.A. Nzeata-Ibe, Ind. J. Phys. (2019) https://doi.org/10.1007/s12648-019-01467-x

[6] C.O. Edet, U.S. Okorie, A.T. Ngiangia and A.N. Ikot, Ind. J. Phys. (2019) https://doi.org/10.1007/s12648-019-01477-9

[7] C.O. Edet and P.O. Okoi, Rev. Mex. Fis. **65,** 333 (2019)

[8] U.S. Okorie, A.N. Ikot, C.O. Edet, I.O. Akpan, R. Sever and R. Rampho, J. Phys. Commun. **3**, 095015 (2019)

[9] C.O. Edet, U.S. Okorie, G. Osobonye, A.N. Ikot, G.J. Rampho and R. Sever, J. Math. Chem. **58**, 989 (2020)

[10] C.O. Edet, P.O. Okoi, A.S. Yusuf, P.O. Ushie and P.O. Amadi, Ind. J. Phys. (2020) https://doi.org/10.1007/s12648-019-01650-0

[11] P.O. Okoi, C.O. Edet and T.O. Magu, Rev. Mex. Fis. **66,** 1 (2020)

[12] C.O. Edet, P.O Okoi, S.O Chima, Rev. Bras. de Ensino de Física, **42**, e20190083 (2020)

[13] A.N. Ikot, U.S. Okorie, A.T. Ngiangia, C.A. Onate, C.O. Edet, I.O. Akpan, P.O. Amadi, Eclética Química J. **45**, 65 (2020)

[14] A.N. Ikot et al. (2020), Submited for publication.

[15] P. Ghosh and D. Nath, Int. J. Quant. Chem. 2020; 1–19. DOI: 10.1002/qua.26153

[16] B.T. Mbadjoun, J. M. Ema'a Ema'a, P. E. Abiama, G. H. Ben-Bolie and P. Owono Ateba Mod. Phys Lett. A **2050092**, 1 (2020)

[17] H. Karayer, Eur. Phys. J. Plus **135**, 70 (2020)

[18] N. Ferkous, A. Bounames, Commun. Theor. Phys. **59**, 679 (2013)

[19] S.M. Ikhdair, B.J. Falaye, , J. Ass. Arab Univ. Basic Appl. Sci. **16,** 1 (2014)

[20] M. Eshghi, H. Mehraban, Eur. Phys. J. Plus **132**, 121 (2017)

[21] M. Eshghi, H. Mehraban, S.M. Ikhdair Eur. Phys. J. A **52**, 201 (2016)

[22] R. Khordad, M.A. Sadeghzadeh, A.M. Jahan-abad, Commun. Theor. Phys. **59**, 655 (2013)

[23] P. Koscik, A. Okopinska, J. Phys. Math. Theor. **40**, 1045 (2007)

[24] M. Aygun, O. Bayrak, I. Boztosun and Y. Sahin, Eur. Phys. J. D **66**, 35 (2012)

[25] M.K. Elsaid, A. Shaer, E. Hjaz and M.H. Yahya, Chin. J. Phys. (2020) https://doi.org/10.1016/j.cjph.2020.01.002

[26] S. Gumber, M. Kumar, M. Gambhir, M. Mohan, and P. K. Jha, Can. J. Phys. **93**, 1 (2015)

[27] A.N. Ikot, U.S. Okorie, G. Osobonye, P.O. Amadi, C.O. Edet, M.J. Sithole, G.J. Rampho and R. Sever. Heliyon **6**, e03738 (2020)

[28] U.S. Okorie, C.O. Edet, A.N. Ikot, G.J. Rampho and R. Sever. Ind. J. Phys. (2020) https://doi.org/10.1007/s12648-019-01670-w

[29] R. Khordad and H.R.R. Sedehi, J. Low Temp. Phys. **190**, 200 (2017)

[30] R. Khordad, Int. J. Thermophys. **34**, 1148 (2013)

[31] R. Khordad, H. Bahramiyan and H.R.R. Sedehi, Opt. Quant. Elect. **50**, 294 (2018)

[32] C.S. Jia, L.H. Zhang, C.W. Wang, Chem. Phys. Lett. **667**, 211 (2017)

[33] C.S. Jia, C.W. Wang, L.H. Zhang, X.L. Peng, R. Zeng, X.T. You, Chem. Phys. Lett. **676**, 150 (2017)

[34] C.S. Jia, X.T. You, J.Y. Liu, L.H. Zhang, X.L. Peng, Y.T. Wang, L.S. Wei, Chem. Phys. Lett. **717**, 16 (2019)

[35] C.S. Jia, L.H. Zhang, X.L. Peng, J.X. Luo, Y.L. Zhao, J.Y. Liu, J.J. Guo, L.D. Tang, Chem. Eng. Sci. **202**, 70 (2019)

[36] R. Jiang, C.S. Jia, Y.Q. Wang, X.L. Peng, L.H. Zhang, Chem. Phys. Lett. **726**, 83 (2019)

[37] A.F. Nikiforov, V.B. Uvarov, Special Functions of Mathematical Physics. Birkhäuser, Basel (1988)

[38] C. Tezcan and R. Sever, Int J Theor Phys **48**, 337 (2009)

[39] S.H. Dong, Factorization Method in Quantum Mechanics, Fundamental Theories in Physics, Vol. 150, Springer, Netherlands (2007)

[40] A. N. Ikot, G. J. Rampho, P. O. Amadi, M. J. Sithole, U. S. Okorie, M.I. Lekala, Eur. Phys. J. Plus **135**, 503 (2020)

[42] M. Eshghi, R. Sever and S.M. Ikhdair, Chin. Phys. B **27,** 020301 (2018)

[42] R.L. Greene and C. Aldrich, Phys. Rev. A **143**, 2363 (1976)

[43] A. Boumali, J Math Chem. **56**, 1656 (2018)

[44] A. Bera, A. Ghosh and M. Ghosh, J. Magnetism and Magnetic Materials **484**, 391 (2019)

[45] M. Hamzavi, S.M. Ikhdair, K.E. Thylwe, J. Math. Chem. **51**, 227 (2013)

[46] J.R. Walton, L.A. Rivera-Rivera, R.R. Lucchese, J.W. Bevan, J. Phys. Chem. A **119** (2016) 6753.

[47] J.L. Hunt, J.D. Poll, L. Wolniewicz,, Can. J. Phys. **62** (1984) 1719.

[48] A.K. Roy, Results Phys. **3** (2013) 103.

[49] J. Baoa, and B. D. Shizgal, Comput. Theoret. Chem. **1149** (2019) 49